\documentclass[reqno, a4paper, 10pt, oneside]{amsart}

\usepackage{amsmath, amssymb, amsthm, mathrsfs, enumerate, scalerel, bbm}
\usepackage[utf8x]{inputenc}
\usepackage[T1]{fontenc}
\usepackage[english]{babel}

\usepackage{geometry}
\usepackage{hyperref}
\hypersetup{colorlinks,linkcolor=blue,citecolor=blue,filecolor=red,urlcolor=blue}

\usepackage[margin = 1cm]{caption}
\usepackage[margin = 0.5cm]{subcaption}
\usepackage{float} %s.t. figures can be placed exactly where they appear in the code
\usepackage{scalerel} %for scaled supstack stuff

\usepackage[sort, nocompress]{cite} %to have the citations appear sorted by number, uncompressed

\makeatletter
\renewcommand{\@biblabel}[1]{\quad#1.}
\makeatother

\usepackage{fancyhdr}
\graphicspath{{plots/}}
\newtheorem{theorem}{Theorem}

\theoremstyle{remark}

\numberwithin{equation}{section}
\numberwithin{theorem}{section}

\newcommand{\E}{\mathbb{E}}
\newcommand{\R}{\mathbb{R}}

\newcommand{\ind}{\text{\ensuremath{1\hspace*{-0.9ex}1}}}
\newcommand{\itembullet}{\,\begin{picture}(-1,1)(-1,-3)\circle*{2}\end{picture}\ }
\title[ ]{Robust Option Pricing with Volatility Term Structure --- An Empirical Study for Variance Options}
\author{Alexander M. G. Cox}
\address{Department of Mathematical Sciences, University of Bath}
\email{a.m.g.cox@bath.ac.uk}

\author{Annemarie M. Grass}
\address{Faculty of Mathematics, University of Vienna}
\email{annemarie.grass@univie.ac.at}
\date{\today.}

\thanks{\textit{Funding:}
A.~Grass gratefully acknowledges financial support through FWF project P 35197 \textit{(Grant DOI: 10.55776/P35197)}}
\begin{document}
\maketitle
%%%%%%%%%%%%%%%%%%%%%%%%%%%%%%%%%%%%%%%%%%%%%%%%%%%%%%%%%%%%%%%%%%%%%%%%%%%%%%%%%
%%%%% ABSTRACT                                                                
%%%%%%%%%%%%%%%%%%%%%%%%%%%%%%%%%%%%%%%%%%%%%%%%%%%%%%%%%%%%%%%%%%%%%%%%%%%%%%%%%

\begin{abstract}
  The robust option pricing problem is to find upper and lower bounds on fair
  prices of financial claims using only the most minimal assumptions.
  It contrasts with the classical, model-based approach and gained
  prominence in the wake of the 2008 financial crisis, and can be used
  to understand the extent to which a model-based price is sensitive
  to the underlying model assumptions. Common approaches involve
  pricing exotic derivatives such as variance options by incorporating
  market data through implied volatility. The existing literature
  focuses largely on incorporating implied volatility information
  corresponding to the maturity of the exotic option. In this paper,
  we aim to explain how intermediate data can and should be
  incorporated.

  It is natural to expect that this additional information will
  improve the robust pricing bounds. To investigate this question, we
  consider variance options, where the bounds of the informed robust
  pricing problem are known.  We proceed to conduct an empirical study
  uncovering a surprising finding: Contrary to common belief, the
  incorporation of more information does not lead to an improvement of
  the robust pricing bounds.

\medskip


\medskip

%\noindent \textit{MSC 2020.} Primary 90C31, 60G65, 60H05; secondary 91G10
\end{abstract}
%
%
%
%%%%%%%%%%%%%%%%%%%%%%%%%%%%%%%%%%%%%%%%%%%%%%%%%%%%%%%%%
\section{Introduction} 
%%%%%%%%%%%%%%%%%%%%%%%%%%%%%%%%%%%%%%%%%%%%%%%%%%%%%%%%%
%
It is well understood by market participants and regulators that the problem of pricing financial derivatives is highly susceptible to model risk 
--- that is, the risk associated with using incorrect pricing models. 
One aspect of reducing model risk is to ensure that pricing models accurately reflect quoted market prices (calibration), however by itself, this is insufficient to remove all model risk, since there are typically multiple models which can fit the same set of market prices. 
It is therefore necessary to understand the set of all models which calibrate to a given set of market prices. By increasing the set of prices that we require a model to calibrate to, we reduce the class of models which can calibrate to the data, and we would therefore expect to get a smaller class of models that are consistent with increased market data, and therefore a better understanding of possible behaviour implied by market prices.

A natural question that arises in practice is: given all relevant prices, can I significantly reduce the range of models which are consistent with observed market prices? In this paper, we will consider this question in the case of market prices for options on variance, given market information in the form of the implied variance term-structure, with information about all available maturities up to the maturity date of the variance option.

Classical approaches to derivative pricing rely on calibrating a class of parametric models to market prices of vanilla options (for example, in the equity context, European Call options). 
In practice, this process leads to inconsistencies, since parametric models are generally insufficient to capture all possible market features, across multiple maturities for both long- and short-dated maturities. 
On the other hand, more flexible models (such as Dupire's Local Volatility model \cite{Du94}) offer greater versatility, but are known not to match market dynamics well, \cite{Ga06}, 
and as a result, are not always reliable models for pricing exotic options.

An alternative approach lies in methods generally called robust, or model-independent pricing. 
In this approach, dating back to the seminal paper of Hobson \cite{Ho98a}, one looks to find prices corresponding to models which are extremal, while imposing as few bounds on the price of options as possible. 
In this context, one is typically able to identify structural properties of extremal models, as well as potential super- or sub-hedging strategies which are robust to model-misspecification. 
Since the original paper of Hobson, a substantial literature has been developed on this topic, see for example, \cite{HeObSpTo16, BrHoRo01a, CoOb11a,  CoOb11b, CoOb09, HoPe02, CoKi19a} among others. 

In the robust setting, one looks for hedging strategies which will work in any market scenario which is possible, 
given the possibility of trading in any liquid options. 
If one increases the set of liquid options that can be used to super-replicate a given exotic option, clearly one has a larger class of possible trading strategies, and therefore a lower model-independent super-replication price. Hence increasing the class of traded options both reduces the minimal super-replication price, and increases the maximal sub-replication price. As a consequence, including more options into the set of liquid options should tighten the bounds on the set of prices which are admissible without (model-independent) arbitrage. By a duality argument (and as a consequence of e.g. \cite{ChKiPrSo21}), one can also see this as a reduction in the set of pricing models which are consistent with observed market prices.

In general, this perspective allows us to understand inclusion of additional liquid options as a method of potentially generating tighter pricing bounds, and therefore getting a tighter constraint on the extent of model-risk inherent in a given price of an exotic option.

In this paper, we consider the case of options on variance, and in particular, variance calls, that is, options which pay the holder $(V_T-K)_+$, where $V$ is the integrated quadratic variation of the log-asset price, that is, if $\sigma_t$ is the observed volatility process of the stock price $(P_t)_{t \in [0,T]}$, then $V_T = \int_0^T \sigma^2_t \, dt$. 
These derivatives are of interest due in part to their close connection to other options on variance/volatility such as the VIX index, and also in the model-independent context because the variance swap $(V_T-K)$, is well known to have a model-free replication strategy (see \cite{DaObRa10}). As a result, we might expect model-independent approaches to be informative about (although not completely fix) prices of, for example, variance calls.

We consider specifically whether the model-independent pricing bounds narrow with the incorporation of additional information contained in European call options with maturity earlier than the maturity date of the variance call options. 
The essence of our approach relies on a construction which allows us to identify the optimal models with the construction of certain space-time boundaries. We are able to informally equate the presence of additional (useful) information of earlier marginals with geometric properties of the underlying boundary structure. 
Moreover, by fitting market data, we are able to compute the space-time boundaries of market data, and conduct an empirical study which indicates that, at least at the times tested, the additional information content coming from earlier marginal information appears to be negligible, indicating that incorporating information contained in the full implied volatility surface is of little benefit for understanding the model risk inherent in variance call options.

One implication of our results is that market participants who look to robustly hedge variance call options using vanilla options to mitigate their risk (see for example \cite{CoOb11b}) will not see substantial benefit through the use of call options from earlier maturities. 
Since model-independent trading strategies of variance call options are known for single maturity options due to \cite{CoWa13}, 
this suggests that these trading strategies are sufficient for many practical situations.

We proceed as follows. We first summarise the model-independent, or robust approach to determining worst case bounds for option prices subject to model risk. 
Then we explain how this connects the case of options on variance to the solutions of Root and Rost of the Skorokhod embedding problem in the single and multiple-marginal settings, and how these solutions can be numerically computed to reveal their geometric structure. 
In Section~\ref{sec:empirical-study}, we describe the market data which we will use to assess our findings. 
In particular, in order to get reliable numerical results for the underlying geometric structure, we need to smooth market prices, which are inherently noisy, as well as renormalisation to identify appropriate estimates for interest rate and dividend parameters. 
Our approach to this relies on calibrating market data to well known models, where we can reliable interpolate complete distributions. 
This is explained in Section~\ref{sec:empiric-results}. We conclude by presenting our numerical results, and showing that the resulting models do not appear to offer any evidence that the multi-marginal setting offers any benefit over calibration for the single marginal setting.

% % % % % % % % % % % % % % % % % % % % % % % % % % % % % %
\subsection{Worst Case Bounds}
% % % % % % % % % % % % % % % % % % % % % % % % % % % % % %
Throughout this article we denote the price process of our underlying asset by $(P_t)_{t \in [0,T]}$. 
We consider a pay-off function $F:C[0,T]\rightarrow \R$ and are interested in determining a fair price $F_T$ of the claim 
\(
    F\left((P_t)_{t \in [0,T]}\right)
\) 
with \emph{maturity} $T>0$.  

Focusing on pricing these derivatives with maximum caution we could consider \emph{all} possible models for 
$(P_t)_{t \in [0,T]}$ satisfying only the most minimal assumptions. 
That is, under a pricing measure a price process is a continuous, non-negative martingale process which agrees with the present-day market value $P_0$. 
Let us denote the set of all such martingales by $\mathcal M^{P_0}$. 
It is important to note here, for the sake of simplicity of this discussion,
we currently assume zero interest rates and dividend yields as otherwise $(P_t)_{t \in [0,T]}$ will only be a martingale after discounting. 
We will, however, give all necessary details for non-zero interest rates and dividend yields in Section \ref{sec:empirical-study}.

\emph{Worst Case Bound} or \emph{robust bound} pricing entails computing an \emph{upper} and \emph{lower} limit of all possible prices.
We then find the following bounds for the price of the derivative $F_T$:
\begin{equation} \label{eq:robust-bounds}
    %\underline F := 
    \inf_{M\in\mathcal M^{P_0}} \E \left[F\left((M_t)_{t \in [0,T]}\right)\right] 
        \leq F_T 
    \leq \sup_{M\in\mathcal M^{P_0}} \E\left[F\left((M_t)_{t \in [0,T]}\right)\right]. 
    %=: \overline F.
\end{equation}
It, however, quickly becomes apparent that this class $\mathcal M^{P_0}$ is too extensive to make much sense of such bounds. 
Numerous authors have suggested and contributed various restrictions of $\mathcal M^{P_0}$ to some sensible subclasses. 
For instance \cite{DaHo07, BoNu13, AcBePeSc13} consider the case where information on a finite number of financial derivatives is available. 
If a continuum of European call options at multiple time points is liquidly traded, market data yields information about the marginal distribution of the price process by the Breeden-Litzenbeger theorem \cite{BrLi78}. The corresponding pricing problem has been studied from a martingale transport perspective, see \cite{HoNe12, GaHeTo13, BeJu16, BeNuTo16, ChKiPrSo21} among many others.
It is well known that under mild technical conditions, the upper limit of prices is equal to the minimal model-independent super-hedging price, and vice versa, the lower limit is equal to the maximal model-independent sub-hedging price, see for example \cite{DoSo14, BeCoHuPePr16}. 

In this article we adopt an approach first proposed by Hobson in \cite{Ho98a} to address this challenge. 
Said method uses market given European call option data to inform on our set of plausible models, 
to then establish connections to specific solutions to the so called \emph{Skorokhod embedding problem} to find concrete optimizer of \eqref{eq:robust-bounds}. 

We give more details in the next section and also refer readers to \cite{Ho11} for an extensive survey. 
Furthermore, we will give an example of a class of claims, namely variance options, which can be examined very rigorously with methods developed in the recent years. 
While incorporating time $T$ market data to refine the set $\mathcal M^{P_0}$ is a well studied problem, 
only recently the necessary theory was developed which allows to extend such results to considering all available European call option data up to time $T$. 

Our aim is to investigate if, in fact, this additional information would indeed results in tighter robust bounds, as one would expect.
In Section \ref{sec:empirical-study} we carry out an in-depth empirical study on variance options using real market data quoted for multiple maturities, 
examining this very question. 
%
%
%
%
%
%
%
%
%
%
%%%%%%%%%%%%%%%%%%%%%%%%%%%%%%%%%%%%%%%%%%%%%%%%%%%%%%%%%
\section{Market-Informed Robust Pricing and the Skorokhod Embedding Problem} \label{sec:robust-pricing}
%%%%%%%%%%%%%%%%%%%%%%%%%%%%%%%%%%%%%%%%%%%%%%%%%%%%%%%%%
%
It is a well-established perspective, as articulated in early works like \cite{Du93, Du94} by Dupire, 
to no longer perceive European call options merely as derivatives of underlying assets to be priced and 
instead to view them as independent assets with their prices given by market dynamics.
Adopting this sentiment we can invoke the Breeden-Litzenberger Theorem \cite{BrLi78}, stating that under the assumption that European call options are computed as the discounted payoff under a pricing measure $\mu$, 
\[
    \mathcal{C}(K,T) = \E_{\mu} \left[ \left(P_T-K\right)^+ \right]
\]
for every strike price $K \in (0, \infty)$ and maturity $T>0$, and furthermore assuming that the pricing function $\mathcal{C}$ is twice differentiable we can recover the pricing measure $\mu$ via
\begin{equation*}
    \mu(P_T \in \mathrm{d} K) = \frac{\partial^2}{\partial K^2}\mathcal{C}(K,T).
\end{equation*}
We can hence refine our set of plausible models $\mathcal M^{P_0}$ to $\mathcal M_{\mu}^{P_0}$, 
the set of those continuous, non-negative martingale models that adhere to the time-$T$ market-given marginal distribution $\mu$, 
precisely
\[
    \mathcal M_{\mu}^{P_0}:= \left\{(M_t)_{t\in [0,T]} \text{ is a continuous, non-negative martingale such that } M_0=P_0 \text{ and } M_T \sim \mu \right\}.
\]

This allows us to refine the uninformed worst case bounds \eqref{eq:robust-bounds} to the following market-informed robust bounds 
\begin{equation} \label{eq:informed-robust-bounds}
    \underline F := 
    \inf_{M\in\mathcal M_{\mu}^{P_0}} \E \left[F\left((M_t)_{t \in [0,T]}\right)\right] 
        \leq F_T 
    \leq \sup_{M\in\mathcal M_{\mu}^{P_0}} \E\left[F\left((M_t)_{t \in [0,T]}\right)\right] 
    =: \overline F.
\end{equation}
Let now $(X_t)_{t \geq 0}$ denote a continuous, non-negative martingale subject to $X_0 = P_0$. 
Consider a stopping time $\tau$ and define the time-changed martingale $(M_t)_{t \geq 0}$ as follows
\begin{equation} \label{eq:martingale}
    M_t := X_{\frac{t}{T - t} \wedge \tau}.
\end{equation}
Note that $M_T = X_{\tau}$, thus we have the equivalence
\begin{equation} \label{eq:model-equiv}
    (M_t)_{t \geq 0} \in \mathcal M_{\mu}^{P_0} \,\, \Leftrightarrow \,\, X_{\tau} \sim \mu. 
\end{equation}
Such stopping times are precisely the solutions to the (generalized) \emph{Skorokhod embedding problem} \eqref{SEP},
where given a process $(X_t)_{t \geq 0}$ and a measure $\mu$ the task is to find a stopping time $\tau$ such that
\begin{equation} \label{SEP}
    X_{\tau} \sim \mu \tag{SEP$_{X, \mu}$}.
\end{equation}
This problem was initially proposed and solved by Skorokhod himself \cite{Sk61, Sk65} in the early 1960s. 
However, it continues to be a subject of extensive research, 
with numerous distinct solutions being established in recent years. 
More than 20 different solutions are surveyed in \cite{Ob04}, 
and substantial contributions have emerged thereafter, notably in 
\cite{CoHo07, CoWa12, CoWa13, GaMiOb14, ObDoGa14, HeObSpTo16, ObSp17, BaBe20, 
       BaZh21, BeCoHu14, BeCoHu16, GhKiPa19, CoObTo19, GaObZo21, BeNuSt19}.
Most of these solutions to \eqref{SEP} distinguish themselves through unique additional extremal properties, 
particular examples will be presented in the next section. 

The application of these solutions to the robust pricing problem \eqref{eq:informed-robust-bounds} has been pioneered by Hobson in \cite{Ho98a}. 
The idea, initially focused on lookback options, is to cleverly select $(X_t)_{t \geq 0}$ and a corresponding specific $\tau$ solutions to \eqref{SEP} such that the bounds in \eqref{eq:informed-robust-bounds} are attained by the martingale $(M_t)_{t \geq 0}$ defined in \eqref{eq:martingale}. 

This idea has been extended in subsequent works, such as \cite{HeObSpTo16}. 
Other types of options have been considered, 
Barrier options were treated in \cite{BrHoRo01a, CoOb11b}, 
forward-start digital options in \cite{HoPe02}, 
double touch/no-touch options in \cite{CoOb11a, CoOb11b} 
and options on leverage exchange traded funds in \cite{CoKi19a, CoKi19b}.

Of particular interest to our study are the extensions to options on variance, as explored in \cite{Ne94, CaLe10, CoWa12, CoWa13, HoKl12} among others. 
Further details on this topic will be provided in the subsequent section.

%
%
%
%
%
%
%
%
%
%
% % % % % % % % % % % % % % % % % % % % % % % % % % % % % %
\subsection{Options on Variance} \label{sec:variance-options}
% % % % % % % % % % % % % % % % % % % % % % % % % % % % % %
%
We now turn our focus to a special class of financial derivatives, namely \emph{variance options}. 
Variance options are contracts written on the quadratic variation of the log price process $\langle \log P\rangle_T$, 
more precisely for some function $f:[0, \infty) \rightarrow \mathbb{R}$, a variance option has a payoff of the form
\[
    f \left(\langle \log P\rangle_T \right).
\]
Frequently considered examples are 
\begin{itemize}
\item[\itembullet] $F = f(\langle \log P\rangle_T ) = \langle \log P\rangle_T - K$, a \emph{variance swap} or
\item[\itembullet] $F = f(\langle \log P\rangle_T ) = (\langle \log P\rangle_T - K)^+$, a \emph{variance call}. 
\end{itemize}
It is worth noting that the quadratic variation of the log price process is invariant under discounting. 

\begin{figure}%[H]
\centering
\begin{subfigure}{.35\linewidth}
\centering
\includegraphics[width = 1\linewidth]{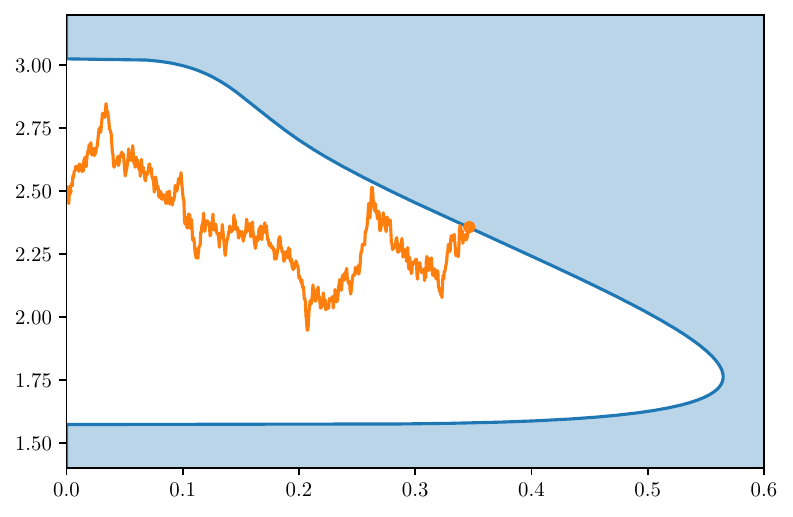}
%\caption{Root solution.}
\end{subfigure}
\quad \quad
\begin{subfigure}{.35\linewidth}
\centering
\includegraphics[width = 1\linewidth]{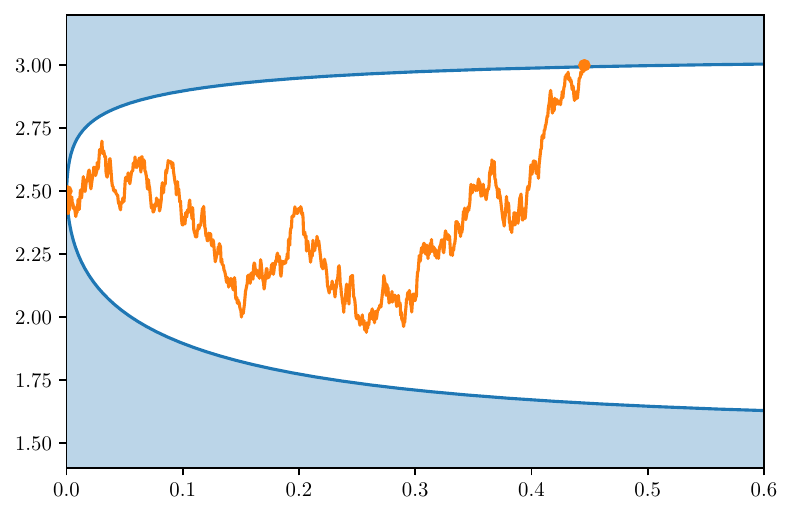}
%\caption{Rost solution.}
\end{subfigure}
\caption{Illustration of a Root (left) resp. Rost (right) solution to \eqref{SEP}.}
\label{fig:Root-Rost-barriers}
\end{figure}

Intriguingly, in \cite{Du05} and more recently in \cite{CaLe10, CoWa12, CoWa13} 
the authors connected the robust pricing problem of variance options to two early and illustrative solutions to \eqref{SEP}, 
namely the Root \cite{Ro69} and Rost \cite{Ro73} solutions. 
We state an existence theorem of these solutions and refer to Figure \ref{fig:Root-Rost-barriers} for a graphical illustration.
\begin{theorem}[\cite{Ro69, Ro73, CoWa12, CoWa13, GaMiOb14, ObDoGa14, GaObZo21}] \label{thm:Root-Rost}
Let $(X_t)_{t \geq 0}$ denote a continuous Markov martingale. 
Consider a measure $\mu$ and an increasing convex function $f:[0, \infty) \rightarrow [0, \infty)$. 
Then
\begin{itemize}
\item[\itembullet]  there exists a \emph{(Root) barrier} $R \subseteq \R^2$ such that
\[    
\tau_{\text{Root}} = \inf \big \{t \geq 0 : (t, X_t) \in R \big \}
\]
is a stopping time solving the \eqref{SEP}.
A \emph{(Root) barrier} is a set $R$ such that  $(t,x) \in  R$ implies $(s,x) \in  R$ for any $s \geq t$.

Moreover $\tau_{\text{Root}}$ \emph{minimizes} $\E[f(\tau)]$ among all solutions $\tau$ to \eqref{SEP}. 
\item[\itembullet]  there exists an \emph{inverse (Rost) barrier} $\bar R \subseteq \R^2$ such that
\[
\tau_{\text{Rost}} = \inf \big \{t \geq 0 : (t, X_t) \in \bar R \big \}
\]
is a stopping time solving the \eqref{SEP}.
An \emph{inverse (Rost) barrier} is a set $\bar R$ such that  $(t,x) \in  \bar R$ implies $(s,x) \in  \bar R$ for any $s \leq t$.

Moreover $\tau_{\text{Rost}}$ \emph{maximizes} $\E[f(\tau)]$ among all solutions $\tau$ to \eqref{SEP}. 
\end{itemize}
\end{theorem}
We will use these two concrete solutions to \eqref{SEP} and use observation \eqref{eq:model-equiv} 
to construct two concrete candidate models contained in $\mathcal M_{\mu}^{P_0}$.

Let $(X_t)_{t \geq 0}$ denote a \emph{geometric Brownian motion} started in $X_0 = P_0$. 
Precisely, let $(B_t)_{t \geq 0}$ denote a Brownian motion, then $X_t$ will satisfy the SDE
\begin{equation} \label{GBM} 
    \mathrm{d} X_t = X_t \mathrm{d}B_t, \quad X_0 = P_0. \tag{GBM}
\end{equation}
It is explained in great detail in \cite[Section 5.2]{Ho11} as well as \cite[Section 1]{CoWa12} and \cite[Section 2]{CoWa13} 
that any candidate price process can be represented as a geometric Brownian motion run at a different speed, 
using a clever time-change argument. 
Conversely, we can construct candidate price processes as time-changed geometric Brownian motions. 
A martingales defined in the following way 
\begin{equation} \label{eq:M-model}
    M_t := X_{\frac{t}{T-t}\wedge \tau}, t \in [0,T]
\end{equation}
where $\tau$ is a solution to \eqref{SEP} satisfies the following 
\begin{itemize}
\item[\itembullet]  $M_0 = P_0$,
\item[\itembullet]  $M_T = X_{\tau} \sim \mu$,
\item[\itembullet]  and most importantly, the quadratic variation of the log price process has the following representation 
$\langle \log M \rangle_T = \tau$.
\end{itemize}

Given the  Root solution $\tau_{\text{Root}}$ (resp. Rost solution $\tau_{\text{Rost}}$) to \eqref{SEP} 
provided by Theorem \ref{thm:Root-Rost} we can define the two candidate prices processes
\begin{itemize}
\item[\itembullet]  $M_t^{\text{Root}} := X_{\frac{t}{T-t}\wedge \tau_{\text{Root}}}$, $t \in [0,T]$, 
\item[\itembullet]  $M_t^{\text{Rost}} := X_{\frac{t}{T-t}\wedge \tau_{\text{Rost}}}$, $t \in [0,T]$,
\end{itemize}
and call them Root and Rost model respectively.
We can now state that the market-informed robust bounds \eqref{eq:informed-robust-bounds} are attained by these models. 
\begin{theorem}[\cite{Ne94, CaLe10, CoWa12, CoWa13}] \label{thm:var-bound-attainment}
For an increasing convex function $f:[0, \infty) \rightarrow [0, \infty)$ consider the variance option $F\left((P_t)_{t \in [0, T]}\right) =  f(\langle \log P \rangle_T)$.
Then for any price $F_T$ of this claim we have the following robust bounds
\begin{equation}
    \E \left[ F \left (M^{\text{Root}}\right) \right] 
    = \inf_{M\in\mathcal M_{\mu}^{P_0}} \E \left[ F \left((M_t)_{t \in [0,T]} \right) \right]
    \leq F_T
    \leq \sup_{M\in\mathcal M_{\mu}^{P_0}} \E \left[ F \left((M_t)_{t \in [0,T]} \right) \right]
    = \E \left[F \left(M^{\text{Rost}} \right) \right]
\end{equation}
More precisely, the lower and upper bound on these options on variance are attained by a Root and Rost model respectively.
\end{theorem}
Given the nice graphical properties of Root and Rost solutions to \eqref{SEP}, 
Theorem \ref{thm:var-bound-attainment} offers a possibility of gaining deeper insights into the behaviour of 
such extremal models attaining the robust pricing bounds. 

While the early works on Root and Rost stopping times were purely existential,  
various more recent works like \cite{CoWa12, CoWa13, GaMiOb14, ObDoGa14, De18} 
offer the necessary tools to analyse such models more concretely and in-depth. 

Incorporating the market-given marginal at the expiration time $T$ is a well studied problem with established results.
But before we apply this theory to market data we aim to generalize 
the results to not only incorporate the time $T$ marginal at expiration, 
but to furthermore incorporate \emph{all} maturities available on the market before said time $T$. 

This extension will be explored in the following section.
%
%
% % % % % % % % % % % % % % % % % % % % % % % % % % % % % %
\subsection{Incorporating Multiple Maturities} \label{sec:MM}
% % % % % % % % % % % % % % % % % % % % % % % % % % % % % %
The market informed robust pricing problem has been well studied for the class of plausible models 
$\mathcal M_{\mu}^{P_0}$, making the asset adhere to the market-informed time $T$ marginal distribution at times of expiration. 

However, European call options are traded liquidly and frequently with 
maturities on numerous stocks and indices expiring several times per week.

Given this rich supply of data it becomes apparent that we are very likely able to use more data for our robust pricing problem than the time $T$ market informed marginals, namely the marginal distributions for all expiries quoted in between. 
Consequently, it is reasonable to assert that our plausible model should conform to all these intermediate marginal constraints, 
thereby tightening the robust bounds on the pricing problem.

To formalize this, let $T_1, \dots, T_n=T$ denote the available maturities leading up to the expiration time $T$. 
Correspondingly, we denote the associated marginals recovered through the Breeden-Litzenberger theorem by $(\mu_1, \dots, \mu_n)$.

Then we can define the set of all plausible martingale models adhering to the market given marginals by
\begin{align*}
        \mathcal M_{(\mu_1, \dots, \mu_n)}^{P_0}:= \{&(M_t)_{t\in [0,T]} \text{ is a continuous, non-negative martingale} 
    \\                                                  &\text{ such that } M_0=P_0 \text{ and } M_k \sim \mu_k, k=1,\dots,n \}
\end{align*}
and consider the refined robust bounds
\begin{equation} \label{eq:MM-informed-robust-bounds}
    \underline F := 
    \inf_{M\in\mathcal M_{(\mu_1, \dots, \mu_n)}^{P_0}} \E \left[F\left((M_t)_{t \in [0,T]}\right)\right] 
        \leq F_T 
    \leq \sup_{M\in\mathcal M_{(\mu_1, \dots, \mu_n)}^{P_0}} \E\left[F\left((M_t)_{t \in [0,T]}\right)\right] 
    =: \overline F.
\end{equation}

Similarly to the single-marginal case, we can again establish a connection between models in 
$\mathcal M_{(\mu_1, \dots, \mu_n)}^{P_0}$ and solutions to a Skorokhod embedding problem. 
An analogous observation to the equivalence \eqref{eq:model-equiv} can be made if instead of the \eqref{SEP} we consider its multi-marginal extension:

For a sequence of measures $(\mu_1, \dots \mu_n)$ and a process $(X_t)_{t \geq 0}$ 
the \emph{multi-marginal Skorokhod embedding problem} \eqref{MMSEP} is to find a sequence of stopping times 
$\tau_1 \leq \cdots \leq \tau_n$ such that
\begin{equation} \label{MMSEP}
    X_{\tau_k} \sim \mu_k \,\,\text{ for }\,\, k = 1, \dots, n \tag{MMSEP$_{X, (\mu_1, \dots \mu_n)}$}.
\end{equation}

Let $(X_t)_{t \geq 0}$ denote a continuous martingale subject to $X_0 = P_0$. 
Given a sequence of market-given marginals $(\mu_1, \dots \mu_n)$, denote by $(\tau_1, \dots , \tau_n)$ a solution to \eqref{MMSEP}. 
With such solution, analogously to the single marginal case we can construct a martingale 
$(M_t)_{t \geq 0} \in \mathcal M_{(\mu_1, \dots, \mu_n)}^{P_0}$ in the following way.
Set $T_0 = \tau_0 = 0$ and define
\begin{equation} \label{eq:MM-model}
    M_t := \sum_{i=1}^{n} X_{\tau_{i-1} + \frac{t-T_{i-1}}{T_{i} - t} \wedge (\tau_{i} - \tau_{i-1}) } \ind_{t \in (T_{i-1}, T_i]} + P_0.
\end{equation}
Then analogously to the single marginal case, 
a model defined in this way will satisfy the following
\begin{itemize}
\item[\itembullet] $M_0 = P_0$,
\item[\itembullet] $M_{T_k} = X_{\tau_k} \sim \mu_k$ for $k = 1, \dots, n$,
\item[\itembullet] $\langle \log M \rangle_T = \tau_n$.
\end{itemize}

\medskip
The existence of multi-marginal Root and Rost solution to \eqref{MMSEP} are ensured by \cite{BeCoHu16, CoObTo19, CoGr23}. 
It is furthermore established in \cite{BeCoHu16} that these solutions still exhibit 
the same extremal properties necessary to identify Root and Rost solutions as optimizers of the right kind of optimization problem. 

More precisely, for an increasing convex function $f:[0, \infty) \rightarrow [0, \infty)$,
the sequence of Root solutions $(\tau^{\text{Root}}_1, \dots ,\tau^{\text{Root}}_n)$ minimizes 
$\E \left[f(\tau_k) \right]$ simultaneously for all $k  = 1, \dots, n$ 
among all other sequences of stopping times $(\tau_1, \dots , \tau_n)$ solving the \eqref{MMSEP}. 
Similarly, a sequence of Rost solutions $(\tau^{\text{Rost}}_1, \dots ,\tau^{\text{Rost}}_n)$ maximizes 
$\E \left[f(\tau_k) \right]$ simultaneously for all $k  = 1, \dots, n$. 

We can now define a multi-marginal Root (resp. Rost) model. 
Let $(X_t)_{t \geq 0}$ denote a geometric Brownian motion \eqref{GBM} started in $X_0 = P_0$. 
Given the multi-marginal Root solution $\tau^{\text{Root}_1}, \dots, \tau^{\text{Root}_n}$ 
(resp. multi-marginal Rost solution $\tau^{\text{Rost}}_1, \dots, \tau^{\text{Rost}}_n$) to \eqref{MMSEP} 
we can utilize Definition \eqref{eq:MM-model} to propose the two candidate prices processes
\begin{itemize}
\item[\itembullet]  $M_t^{\text{Root}} := \sum_{i=1}^{n} X_{\tau^{\text{Root}}_{i-1} + \frac{t-T_{i-1}}{T_{i} - t} \wedge (\tau^{\text{Root}}_{i}-\tau^{\text{Root}}_{i-1})} \ind_{t \in (T_{i-1}, T_i]} + P_0$, $t \in [0,T]$, 
\item[\itembullet]  $M_t^{\text{Rost}} := \sum_{i=1}^{n} X_{\tau^{\text{Rost}}_{i-1} + \frac{t-T_{i-1}}{T_{i} - t} \wedge (\tau^{\text{Rost}}_{i}-\tau^{\text{Root}}_{i-1})} \ind_{t \in (T_{i-1}, T_i]} + P_0$, $t \in [0,T]$,
\end{itemize}
and call them multi-marginal Root and multi-marginal Rost model respectively.

\medskip
For an increasing convex function $f:[0, \infty) \rightarrow [0, \infty)$ let us now consider the variance option 
$F\left((P_t)_{t \in [0, T]}\right) =  f(\langle \log P \rangle_T)$.
Then any fair price $F_T$ of this claim will adhere to the following robust bounds 
informed by the market-given marginals $(\mu_1, \dots, \mu_n)$ as follows
\begin{align}
\begin{split} \label{eq:Root-Rost-bounds}
    \E \left[ F \left (M^{\text{MM-Root}}\right) \right] 
    &=    \inf_{M\in\mathcal M_{(\mu_1, \dots, \mu_n)}^{P_0}} \E \left[ F \left((M_t)_{t \in [0,T]} \right) \right]
\\  \leq F_T
    &\leq \sup_{M\in\mathcal M_{(\mu_1, \dots, \mu_n)}^{P_0}} \E \left[ F \left((M_t)_{t \in [0,T]} \right) \right]
    = \E \left[F \left(M^{\text{MM-Rost}} \right) \right].
\end{split}
\end{align}
It is noteworthy that the inclusion $\mathcal M_{(\mu_1, \dots, \mu_n)}^{P_0} \subseteq \mathcal M_{\mu}^{P_0}$ 
implies that we have  
$\E \left[ F \left (M^{\text{MM-Root}}\right) \right] \geq \E \left[ F \left (M^{\text{Root}}\right) \right]$ 
as well as
$\E \left[ F \left (M^{\text{MM-Rost}}\right) \right] \leq \E \left[ F \left (M^{\text{Rost}}\right) \right]$.
 
Therefore, incorporating more market given marginal into the robust pricing problem should possibly refine the bounds.

\medskip
In the next section we summarize the results required for numerical examination 
of such extremal models achieving the robust pricing bounds informed by the market-given marginals $(\mu_1, \dots, \mu_n)$.
%
%
%
%
%
%
%
%
%
%
% % % % % % % % % % % % % % % % % % % % % % % % % % % % % %
\subsection{Analysis of Root and Rost Models} \label{sec:Root-Rost}
% % % % % % % % % % % % % % % % % % % % % % % % % % % % % %
%
We come to the main objective of this article, to investigate the behaviour of those models 
for which the bounds of the robust pricing problem \eqref{eq:MM-informed-robust-bounds}, incorporating multiple market given marginals are attained. 

These models, namely the multi-marginal Root and Rost models defined in Section \ref{sec:MM} 
are characterized by the respective sequences of barriers which need to be passed by the candidate price processes consecutively. 
Hence, we believe that analysing these barriers might offer valuable insights into the underlying models.

While the original results of Root \cite{Ro69} and Rost \cite{Ro76} are purely of existential nature 
and concrete barriers were not known save for a few very simple examples, 
these solutions later became computationally tractable thanks to \cite{CoWa12, CoWa13, GaMiOb14, ObDoGa14, De18} among others. 

In this single-marginal case, where the robust bounds are only informed by the time $T$ 
marginal distribution at time of expiration as outlined in Section \ref{sec:variance-options}, %in \eqref{eq:informed-robust-bounds}
a thorough numerical investigation of the Root and Rost models was conducted in \cite{CoWa13} using simulated market data. 
Below, we extend these results to the multi-marginal case.
In order to perform a similar meaningful analysis of the multi-marginal Root and Rost models attaining 
the bounds of the robust pricing problem \eqref{eq:MM-informed-robust-bounds} 
informed by multiple maturities, tractable representations of multi-marginal Root and Rost solutions are essential. 
A computable multi-marginal Root solution is provided in \cite{CoObTo19}, 
while a computable multi-marginal Rost solution, to the best of our knowledge, 
has only been developed recently in \cite{CoGr23}.

We will give a brief summary of the well-established single-marginal results 
and proceed to state their extensions to multiple marginals. 

% % % % % % % % % % % % % % % % % % % % % % % % % % % % % %
\subsection*{Single-Marginal Results}
% % % % % % % % % % % % % % % % % % % % % % % % % % % % % %
Let $(X_t)_{t \geq 0}$ represent a geometric Brownian motion \eqref{GBM}. 
Given a measure $\mu$ let $\tau_{\text{Root}}$ be a Root solution to \eqref{SEP}, 
induced by the Root barrier $R$. 
Throughout this paper we consistently assume the process $(X_t)_{t \geq 0}$ 
to start according to the starting measure $\lambda = \delta_{P_0}$, 
however more general starting measure can be considered. 
We refer to the original papers for details.

Define the function $u(t,x) := - \E \left[|X_{\tau_{\text{Root}} \wedge t} - e^x |\right]$, 
and for a measure $m$, let $U_m(x):= \int |x-y|\mathrm{d}m(y)$ denote its \emph{potential}. 
The Root barrier $R$ now also has the following representation
\begin{equation} \label{eq:Root-barrier}
    R = \left\{(t,e^x) : u(t,x) = U_{\mu}(e^x) \right\}.
\end{equation}
However, it is crucial to know and the main result in \cite{CoWa12} that the function $u(t,x)$ can be recovered without any knowledge of $\tau_{\text{Root}}$ 
and $R$ as a solution to the following PDE in variational form
\begin{equation}
\begin{gathered} \label{eq:Root-PDE}
        \max\left\{
                \frac{1}{2}\frac{\partial^2 u}{\partial x^2}(t,x) 
                    - \frac{1}{2}\frac{\partial u}{\partial x}(t,x) 
                    - \frac{\partial u}{\partial t}(t,x), 
                 U_{\mu}(e^x)  - u(t,x)
            \right\} = 0,
    \\ u(0,x) = U_{\lambda}(e^x).
\end{gathered}
\end{equation}
This gives a concrete method for the numerically computing Root barriers by solving Equation \eqref{eq:Root-PDE} 
and recovering the Root barrier $R$ via Equation \eqref{eq:Root-barrier}.

Similarly, let $\tau_{\text{Rost}}$ denote a Rost solution to \eqref{SEP}, induced by the Rost barrier $\bar R$ 
and consider the function $v(t,x) := U_{\mu}(e^x) + \E \left[|X_{\tau_{\text{Rost}} \wedge t} - e^x |\right]$. 
Then the Rost barrier $\bar R$ also has the following representation
\begin{equation} \label{eq:Rost-barrier}
    \bar R = \left\{(t,e^x) : v(t,x) = U_{\mu}(e^x) -  U_{\lambda}(e^x) \right\}.
\end{equation}
Moreover, similar to the Root case and as discovered in \cite{CoWa13, De18} the function $v(t,x)$ 
can also be recovered as the solution to the following PDE in variational form
\begin{equation}
\begin{gathered} \label{eq:Rost-PDE}
        \max\left\{
                \frac{1}{2}\frac{\partial^2 u}{\partial x^2}(t,x) 
                    - \frac{1}{2}\frac{\partial u}{\partial x}(t,x) 
                    - \frac{\partial u}{\partial t}(t,x), 
                 (U_{\mu} - U_{\lambda})(e^x) - v(t,x)
            \right\} = 0,
    \\ v(0,x) = (U_{\mu} - U_{\lambda})(e^x).
\end{gathered}
\end{equation}
%
%
%
% % % % % % % % % % % % % % % % % % % % % % % % % % % % % %
\subsection*{Multi-Marginal Results}
% % % % % % % % % % % % % % % % % % % % % % % % % % % % % %
The preceding results allow for analysis of Root and Rost models associated to options on variance 
when considering the market given marginal at time of maturity $T$. 
However, extending this analysis to a multi-marginal context, to the best of our knowledge, has remained unexplored.

In Section \ref{sec:MM} it was explained how the bounds of the robust pricing problem 
informed by multiple maturities as formulated in \eqref{eq:MM-informed-robust-bounds} 
are attained by a multi-marginal Root and Rost model respectively. 
In order to perform a similarly meaningful investigation of such models as in the single-marginal 
case it remains to establish a method of numerical analysis analogous to the above.  
For the multi-marginal Root case a computable representation is know due to \cite{CoObTo19}, 
while \cite{CoGr23} recently introduced a similar representation for multi-marginal Rost solutions. 
Both these works give an optimal stopping characterization of multi-marginal Root and Rost solutions and state results for embedding into a Brownian motion. 

It remains to discuss the application of these results to a geometric Brownian motion \eqref{GBM} 
and how to establish at a PDE representation similar to \eqref{eq:Root-PDE} and \eqref{eq:Rost-PDE}. 
The process of deriving a PDE representation in variational form starting from an optimal stopping problem is 
well known due to \cite{BeLi82}, we also refer to Section 5.2 in \cite{Ph09} 
for a more accessible explanation. 

To extend such results form a Brownian motion to a geometric Brownian motion we utilize the same methods as described 
in \cite[Section 4.3]{CoWa12} where the authors propose the change of variables $(t,x) \mapsto (t, e^x)$. 
While the transitions densities of a Brownian motion must satisfy the heat equation, 
the transition densities of the \eqref{GBM} will in turn, 
after said change of variables satisfy a drift-diffusion equation as seen on the left hand side 
of the PDEs in variational form \eqref{eq:Root-PDE}, \eqref{eq:Rost-PDE} \eqref{eq:MM-Root-PDE} and \eqref{eq:MM-Rost-PDE}. 
For more details we refer to the proof of \cite[Theorem 4.6]{CoWa12} and the preceding discussion therein. 

Altogether, this leads us to the following characterizations of multi-marginal Root and Rost solutions to \eqref{MMSEP} as solutions to PDEs in variational form. 

Consider now a geometric Brownian motion \eqref{GBM} $(X_t)_{t \geq 0}$ 
and a sequence of measures $\mu_1, \dots, \mu_n$. 

We first consider the multi-marginal Root case, 
hence let $(\tau^{\text{Root}}_1, \dots ,\tau^{\text{Root}}_n)$ 
be a multi-marginal Root solution to \eqref{MMSEP} induced by the sequence of Root barriers $R_1, \dots, R_n$.

For $k = 1, \dots, n$ define the function $u_k(t,x) = - \E \left[|X_{\tau^{\text{Root}}_k \wedge t} - e^x |\right]$ 
as well as $u_0(t,x) = U_{\lambda}(e^x) = U_{\delta_{P_0}}(e^x)$. 

Then the $k$-th Root barrier $R_k$ corresponding to the $k$-th Root stopping time $\tau^{\text{Root}}_k$ has the following representation 
\[
    R_k = \left\{(t,e^x) : u_{k-1}(t,x) - u_k(t,x)  =  \left(U_{\mu_{k-1}} - U_{\mu_k}\right)(e^x)  \right\}.
\]
Similar to the single-marginal case, the functions $u_k(t,x)$ for $k = 1, \dots, n$ 
can be recovered inductively as the solution of the following PDE in variational form
\begin{equation}
\begin{gathered} \label{eq:MM-Root-PDE}
        \max\left\{
                \frac{1}{2}\frac{\partial^2 u_k}{\partial x^2}(t,x) 
                    - \frac{1}{2}\frac{\partial u_k}{\partial x}(t,x) 
                    - \frac{\partial u_k}{\partial t}(t,x), \,
                u_{k-1}(t,x) - u_k(t,x) - \left(U_{\mu_{k-1}} - U_{\mu_k}\right)(e^x)  
            \right\} = 0,
    \\ u_k(0,x) = U_{\lambda}(e^x).
\end{gathered}
\end{equation}
A similar result holds in the multi-marginal Rost case. 
For $k = 1, \dots, n$ define the function $v_k(t,x) = U_{\mu_k}(e^x) - \E \left[|X_{\tau^{\text{Rost}}_k \wedge t} - e^x | \right]$ 
as well as $v_0(t,x) = (U_{\mu_k} - U_{\lambda})(e^x) = (U_{\mu_k} - U_{\delta_{P_0}})(e^x)$.
Then the $k$-th Rost barrier $\bar R_k$ corresponding to the $k$-th Rost stopping time $\tau^{\text{Rost}}_k$ also has the following representation
\[
    \bar R_k = \left\{(t,e^x) :v_{k-1}(t,x) - v_k(t,x) =  \left(U_{\mu_{k-1}} - U_{\mu_k}\right)(e^x)  \right\}.
\]
Again, the functions $v_k(t,x)$ for $k = 1, \dots, n$ can be recovered inductively 
as the solution to the following PDE in variational form
\begin{equation}
\begin{gathered} \label{eq:MM-Rost-PDE}
        \max\left\{
                \frac{1}{2}\frac{\partial^2 v_k}{\partial x^2}(t,x) 
                    - \frac{1}{2}\frac{\partial v_k}{\partial x}(t,x) 
                    - \frac{\partial u}{\partial t}(t,x), \,
                 v_{k-1}(t,x) - v_k(t,x) - \left(U_{\mu_{k-1}} - U_{\mu_k}\right)(e^x) 
            \right\} = 0,
    \\ v_k(0,x) = (U_{\mu_k} - U_{\lambda})(e^x).
\end{gathered}
\end{equation}
Hence, provided market-given marginal $(\mu_1, \dots, \mu_n)$ the above allows us to compute the concrete barriers 
the Root and Rost models must adhere to.
As these barriers each have to be passed through consecutively, 
this possibly gives us valuable insight into the behaviour of these extremal models. 
\subsection*{Computation of robust bounds}
We conclude this section by explaining the computation of prices for variance options given Root and Rost barriers. 
For a function $f:[0, \infty) \rightarrow [0, \infty)$ we consider the variance option 
$F\left((M_t)_{t \in [0, T]}\right) =  f(\langle \log M \rangle_T)$.

For a Root barrier $R$ we define a corresponding \emph{barrier function} $R(x)$ as follows 
\[
 R(x):= \inf \{t \geq 0:(t,x) \in R \},
\]
while for a Rost barrier $\bar R$ such a function will be defined via
\[
    \bar R(x):= \sup \{t \geq 0:(t,x) \in \bar R\}.
\]
Let us now assume that $\mu$ is atomless and the barrier functions $R(x)$ and $\bar R(x)$ are well defined and continuous.   
In this case, we particularly note that
\[
\tau = R(X_{\tau}).
\]
Recall that for our martingale model $(M_t)_{t \in [0,T]}$ defined via \eqref{eq:M-model} or \eqref{eq:MM-model} we have
$\langle \log M \rangle_T = \tau$
where $\tau$ denotes the Root (resp. Rost) barrier corresponding to the terminal maturity $T$. 

Thus, under such a Root (resp. Rost) model, a fair price $F_T$ of our variance option can be easily computed as 
\begin{equation} \label{eq:price}
    %F_T = \mathbb{E}\left[f(\tau) \right] = \int f(R(x)) \varphi_{\mu}(x) \mathrm{d}x. 
    F_T = \mathbb{E}\left[f(\tau) \right] = \int f(R(x)) \mathrm{d} \mu_T(x). 
\end{equation}

We carry out an empirical study on real market data in the next section.

\section{Empirical Study} \label{sec:empirical-study}
%%%%%%%%%%%%%%%%%%%%%%%%%%%%%%%%%%%%%%%%%%%%%%%%%%%%%%%%%
%
%
%
\begin{figure}%[H]
\centering
\includegraphics[width = 1\linewidth]{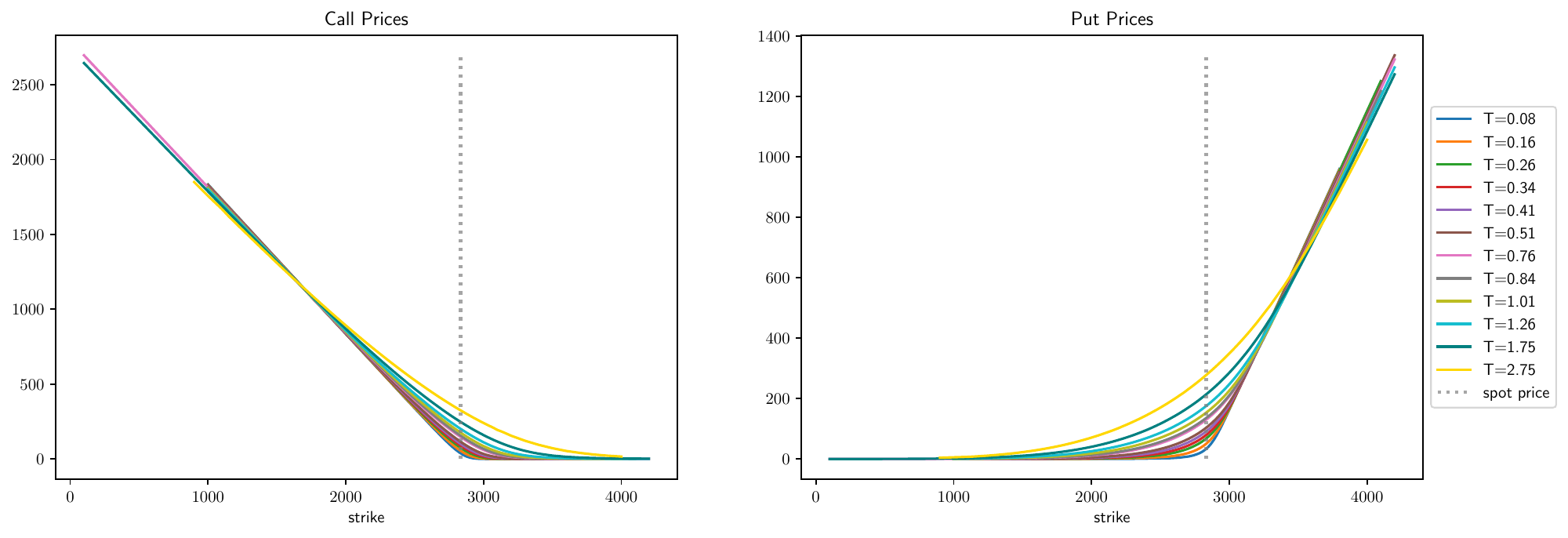}
\caption{Market given European call and put option prices on the S\&P500.}
\label{fig:option-prices}
\end{figure}
We conduct an empirical study utilizing European put and call options data quoted on the S\&P500 index as of March 19, 2019, acquired from the CBOE Datashop. 

We opt for using closing quotes from the dataset for our study, investigating the 12 available (non-weekly) expiries which mature every third Friday of the month.

The shortest expiry within our dataset is 31 days to maturity (0.085 years to maturity), while the longest maturity extends to 1005 days (2.75 years to maturity). 
Furthermore, there are between 83 and 296 prices quoted per maturity, ranging between 100\$ and 4200\$ in strike value while the closing spot price of the S\&P500 was provided by CBOE as $P_0 = 2833\$$. 

The prices sourced from CBOE are provided as bid-ask quotes, for our further analysis we compute and work with the mid-point prices.
These quoted (mid-point) call and put options prices are illustrated in Figure \ref{fig:option-prices}.

While the exact interest rates and dividend yield used for option pricing were not disclosed by CBOE, 
we estimate a constant implied interest rate, denoted as $r$, at $r = 0.0265$, 
and a constant implied dividend yield, denoted as $q$, at $q = 0.0185$. 
Details on our estimation methods can be found in Appendix \ref{A:r-and-q}. 
It is noteworthy, however, that these estimated values are in alignment with the T-bill rates quoted by the U.S. Department of the Treasury, as well as reported annual dividend yields of the S\&P500, promoting confidence in the validity of our empirical setup.

In Section \ref{sec:empiric-preparation} we take the necessary preparations in order to compute 
multi-marginal Root and Rost barriers as detailed in Section \ref{sec:Root-Rost}. 
The numerical results will be presented in Section \ref{sec:empiric-results}. 
%
%
%
%
%
%
%
%
%
%
% % % % % % % % % % % % % % % % % % % % % % % % % % % % % %
\subsection{Preparation of the data} \label{sec:empiric-preparation}
\begin{figure}[H]
\centering
\includegraphics[width = 1\linewidth]{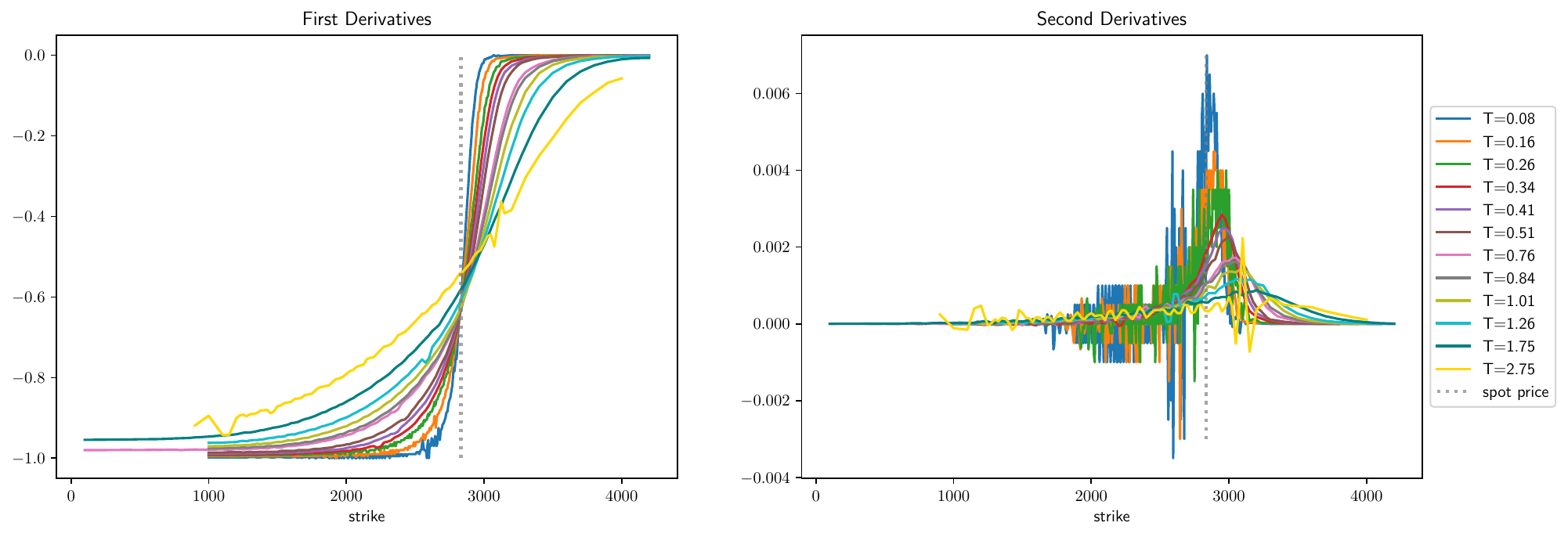}
\caption{First and second derivatives of market given call prices.}
\label{fig:option-derivatives}
\end{figure}

% % % % % % % % % % % % % % % % % % % % % % % % % % % % % %
Market-quoted data is inherently noisy, 
occasionally disrupted by reporting delays or inaccuracies. 
These irregularities can introduce problems like artificial arbitrage opportunities within the data, which may not accurately represent real market conditions. 
Moreover, while the prices themselves may appear orderly, 
closer examination reveals increasing levels of noise in their first and second derivatives, 
as depicted in Figure \ref{fig:option-derivatives}. 

In the context of numerical solutions to partial differential equation, as introduced in Section \ref{sec:Root-Rost},  
these noisy derivatives pose a great challenge, usually resulting in failure of the algorithm. 

There are several possibilities to tackling this problem, 
we choose to investigate two possible calibration scenarios, 
both calibrating a Black-Scholes model and a Heston model to the market given data.
Our rationale for this choice is as follows: 
Despite its recognized limitations, the Black-Scholes model remains a widely used method 
for pricing European call options by numerous financial institutions. 

The Heston model \cite{He93} is know to overcome 
some of the limitations of the Black-Scholes model while remaining tractable, 
hence being a favored alternative to the Black-Scholes model by practitioners. 
Heston models can in practice demonstrate exact fits to market data 
in part due to its widespread use by market participators. 

In sum, rather than considering market prices as direct indicators, 
we view the market as providing the underlying pricing parameters for widely used pricing models.  

This approach is not necessary, provided alternative approaches to smoothing the market data can be used. 
Other potential methods include \cite{DeHe23}
or arbitrage free spline smoothing as suggested in \cite{Fe09}, 
but we have not implemented these methods to date. 

\subsection*{Forward prices.}
Our dataset provides European call option quoted on the spot price process $(P_t)_{t \geq 0}$.
However, as presented in Section \ref{sec:robust-pricing}, it is essential to consider call option prices quoted on the forward price process.
Therefore, the final step in our data preparation requires appropriate modification of the quotes. 

Given a strike price $x$ and a maturity $T$ we will by $\mathcal{C}(x, T)$ denote the price of a European call option on the asset $(P_t)_{t \geq 0}$. 
We furthermore assume this price to be given under a risk neutral (martingale) measure via
\begin{equation*}
    %C(x) = 
    \mathcal{C}(x,T) = e^{-rT} \E \left[ \left(P_T-x\right)^+ \right] %= \E[(S_T - e^{-rT}x)^+] = c(e^{-rT}x)
\end{equation*}
where $r$ denotes the constant and deterministic risk-free interest rate.

As the S\&P500 is a stock index consisting of 500 different stocks, 
each paying dividends at individual times throughout the year we choose 
to model a continuously paid dividend which can effectively be seen as an average of 
all the different dividends paid by the different companies at various times.

Assuming such a continuously compounded dividend yield $q$, we consider the forward price process $(S_t)_{t \geq 0}$
\begin{equation*}
    S_t := e^{(q-r)t} P_t,
\end{equation*}
which is a martingale under the pricing measure. 

By $c(x,T)$ we will denote the time-zero price of a call option on the discounted process $(S_t)_{t \geq 0 }$ with strike $x$ and maturity $T$, 
\[
    c(x,T) := \E \left [(S_T - x)^+ \right].
\]
We then observe the following relationship between $\mathcal{C}(x,T)$ and $c(x,T)$
\begin{equation} \label{eq:discounted-prices}
    c(x,T) = \E \left[ \left(e^{(q-r)T} P_T - x \right)^+ \right] 
         = e^{qT} \left( e^{-rT}\E \left[ \left(P_T - e^{-(q-r)T} x \right)^+ \right] \right)
         = e^{qT}\mathcal{C}\left( e^{-(q-r)T} x, T\right).
\end{equation}
Hence, we use Equation \eqref{eq:discounted-prices} to compute prices quoted on the forward price process given a pricing function $\mathcal{C}$. 

The forward prices associated to our market given data can be seen in Figure \ref{fig:BS} (c) resp. Figure \ref{fig:Heston} (c)
using a pricing function obtained via calibration of a Black-Scholes resp. Heston pricing model to the market given data. 

As presented in Section \ref{sec:Root-Rost}, the necessary input for the computation of Root and Rost barriers are the potentials $U_{\mu^k}(x)$ corresponding to market implied measures $\mu^k$, $k \in \{1, \dots, n\}$. 
Note that the potential of a measure can easily be recovered from the call prices \eqref{eq:discounted-prices} in the following way
\[
    U_{\mu^k}(x) = P_0 - 2c(x, T_k) - x,
\]
we show the potentials for our data in Figure \ref{fig:BS} (c) resp. Figure \ref{fig:Heston} (c).

%
%
%
%
%
%
%
%
%
%
% % % % % % % % % % % % % % % % % % % % % % % % % % % % % %
\subsection{Numerical Results} \label{sec:empiric-results}
% % % % % % % % % % % % % % % % % % % % % % % % % % % % % %
%
In this section we calibrate both a Black-Scholes and a Heston model to our market-given data. 
This calibration process enables us to use the respective call prices as a smoothed approximation 
of our market prices to inform on the robust pricing problem \eqref{eq:MM-informed-robust-bounds}. 

The effectiveness of the respective calibration can be seen in Figure \ref{fig:BS} (a) and Figure \ref{fig:Heston} (a)
where we show a comparison of the calibrated prices to the market-given prices. 
Additionally, we observe the indeed very well-behaved first and second derivatives 
of the calibrated prices in Figure \ref{fig:BS} (b) and Figure \ref{fig:Heston} (b).

\begin{figure}%[H]
\centering
\begin{subfigure}{.45\linewidth}
\centering
\includegraphics[width = 1\linewidth]{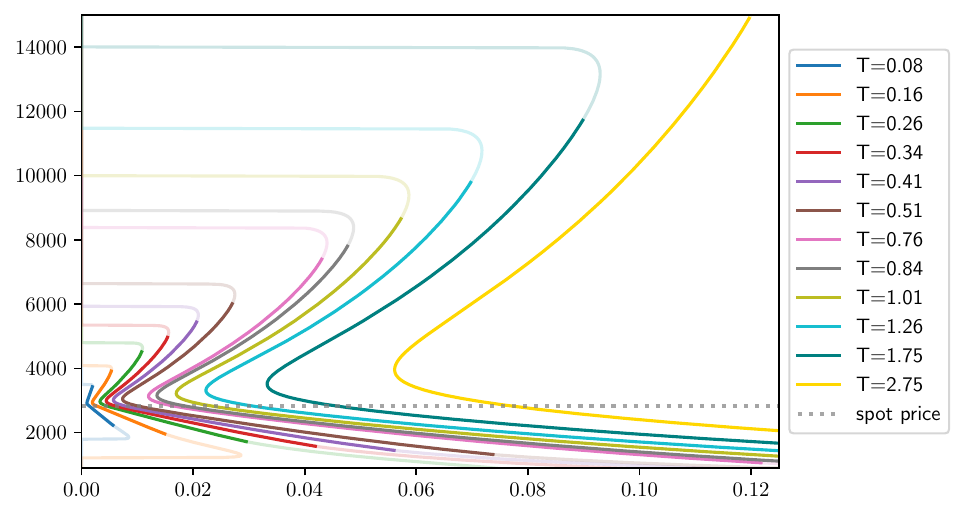}
\caption{Single marginal barriers}
\end{subfigure}
\quad
\begin{subfigure}{.45\linewidth}
\centering
\includegraphics[width = 1\linewidth]{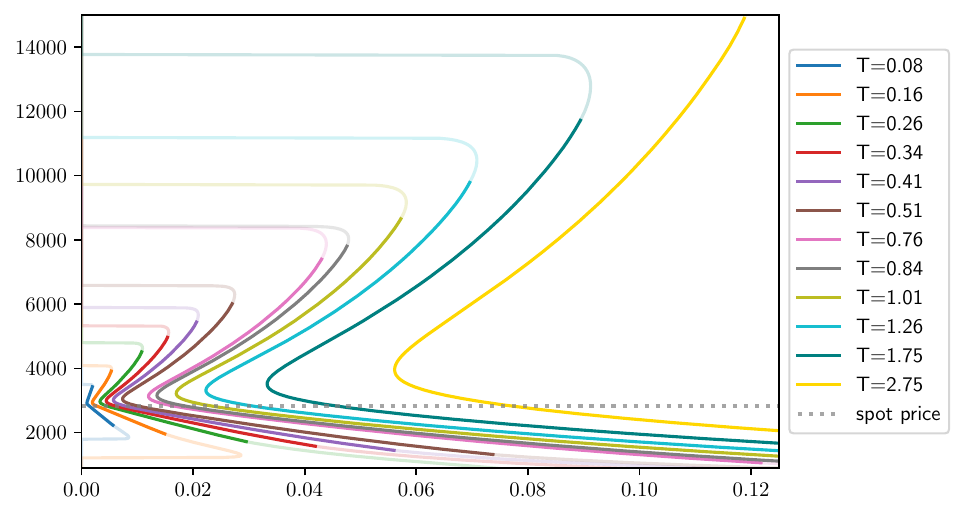}
\caption{Multi marginal barriers}
\end{subfigure}
\centering
\begin{subfigure}{.45\linewidth}
\centering
\includegraphics[width = 1\linewidth]{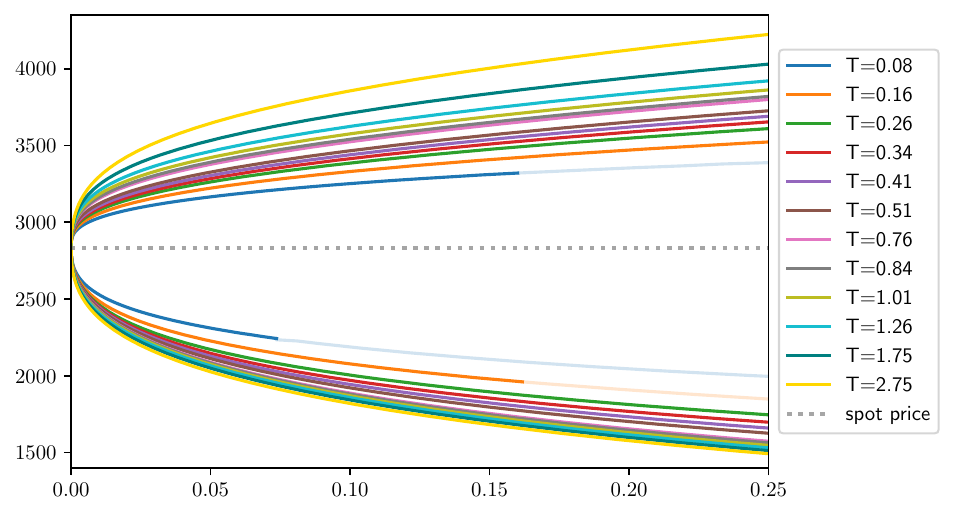}
\caption{Single marginal barriers}
\end{subfigure}
\quad
\begin{subfigure}{.45\linewidth}
\centering
\includegraphics[width = 1\linewidth]{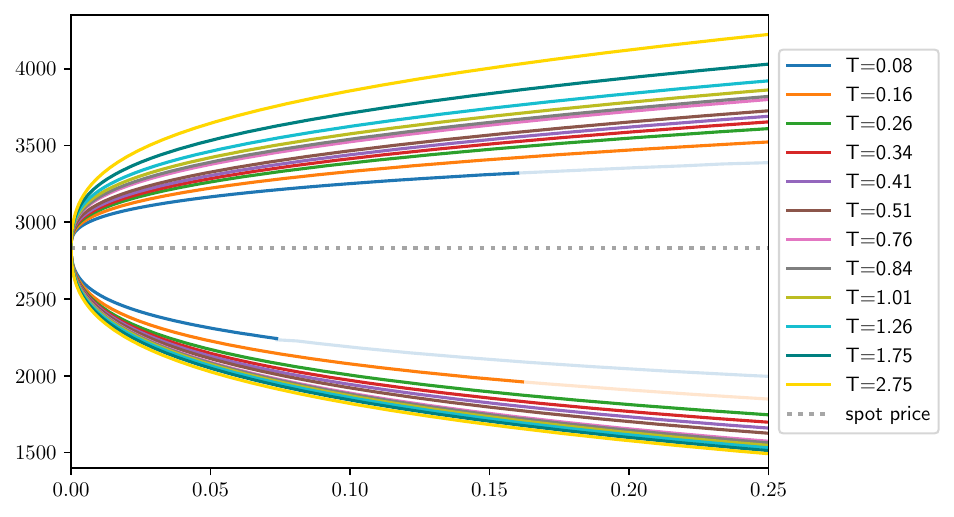}
\caption{Multi marginal barriers}
\end{subfigure}
\caption{Root (top) and Rost (bottom) barriers embedding the market implied marginals of the S\&P500.}
\label{fig:Heston-barriers}
\end{figure}

Using a classic Euler numeric differentiation scheme we first solve the single-marginal embedding problem given 
in \eqref{eq:Root-PDE} resp. \eqref{eq:Rost-PDE} for each maturity. 
Here the stopping time embedding the $k$-th marginal is not required to wait for the stopping time embedding the $(k-1)$-th marginal. 
The results are depicted on the left side in Figure \ref{fig:BS-barriers} (a) and (c) resp. Figure \ref{fig:Heston-barriers} (a) and (c).
On the right these same figures show the numerical solution to the multi-marginal problem given 
in \eqref{eq:MM-Root-PDE} resp. \eqref{eq:MM-Rost-PDE}, namely
in Figure \ref{fig:BS-barriers} (b) and (d) resp. Figure \ref{fig:Heston-barriers} (b) and (d).

We conduct a more extensive discussion of the resulting barriers in Section \ref{sec:discussion}.

%
%
%
%
%
%
%
%
%
%
%
%
% % % % % % % % % % % % % % % % % % % % % % % % % % % % % %  
\subsubsection{Black-Scholes Data} \label{sec:BS}
% % % % % % % % % % % % % % % % % % % % % % % % % % % % % % 
%
\

\noindent
\textbf{Calibration Results.} 
Calibrating a \emph{Black-Scholes model} to the market data %as outlined in the Appendix \ref{A:BS-calibration} 
yields the pricing parameter $\sigma = 0.153$. 

\begin{figure}%[H]
\centering
\begin{subfigure}{0.9\linewidth}
\centering
\includegraphics[width = 1\linewidth]{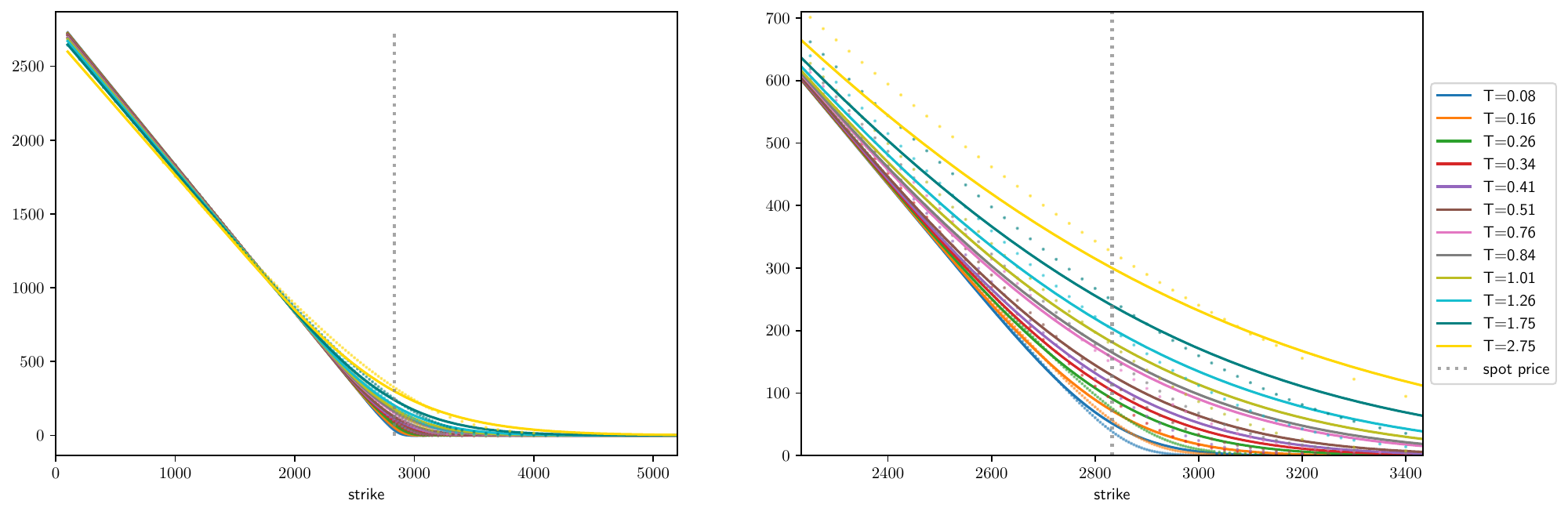}
\caption{Comparing calibrated prices to market given prices.}
\end{subfigure}
\begin{subfigure}{0.9\linewidth}
\centering
\includegraphics[width = 1\linewidth]{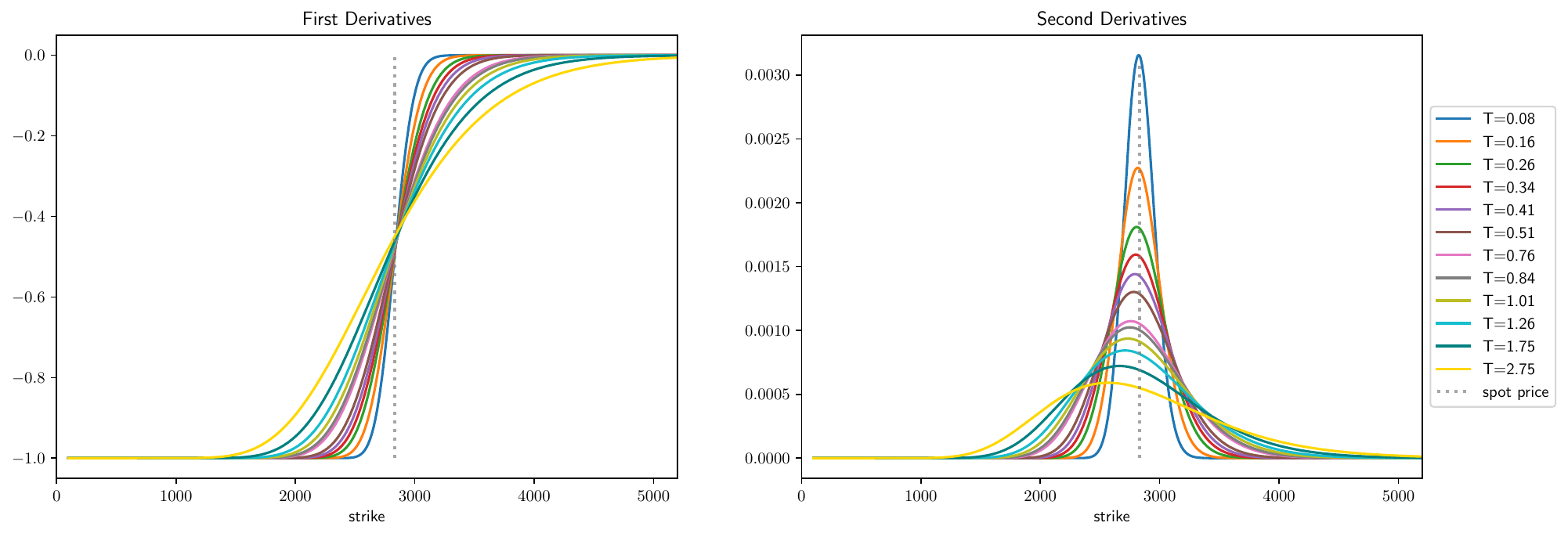}
\caption{First and second derivatives of calibrated prices.}
\end{subfigure}
\centering
\begin{subfigure}{0.9\linewidth}
\centering
\includegraphics[width = 1\linewidth]{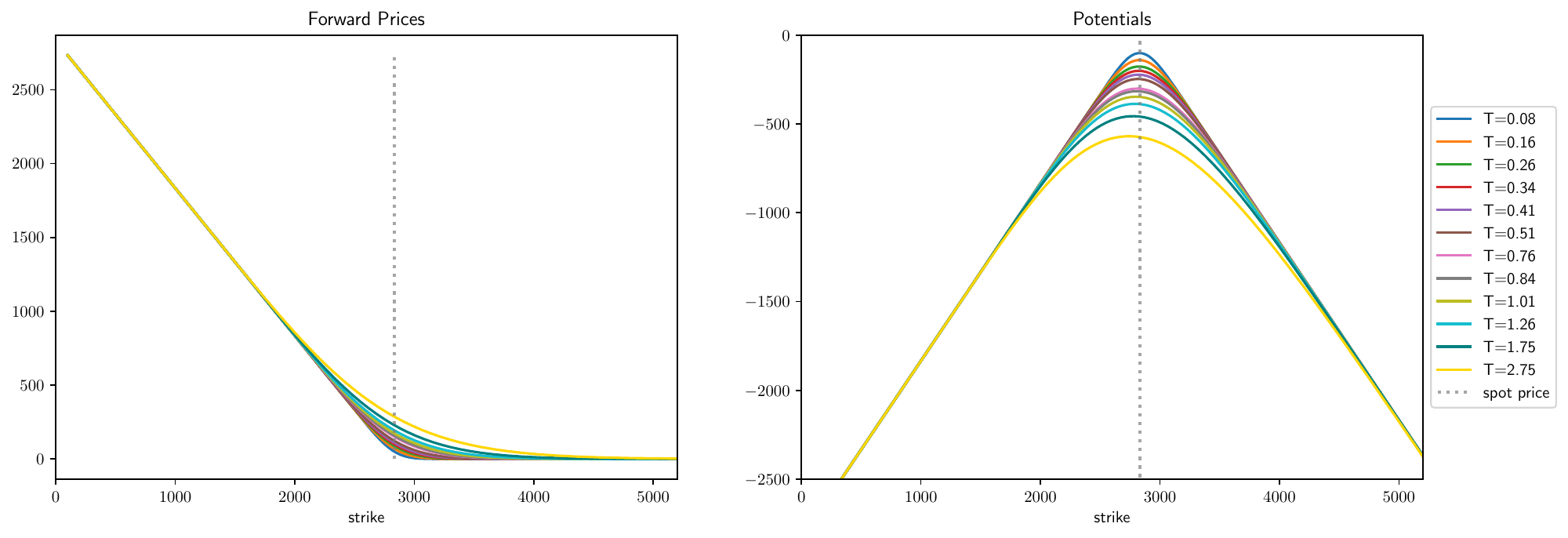}
\caption{Forward prices and the associated potentials.}
\end{subfigure}
\caption{Black-Scholes model calibration results.}
\label{fig:BS}
\end{figure}

We want to point out that the fit of the calibrated model to the market given data as seen in Figure \ref{fig:BS} (a) seems unfavourable and not ideal. 
The Black-Scholes model does however provide very smooth derivatives, depicted in Figure \ref{fig:BS} (b) that allow for easy application of the Root and Rost barrier computation algorithm. 
The Black-Scholes forward prices and the associated potential functions used for the algorithm can be seen in Figure \ref{fig:BS} (c). 
\subsection*{Root and Rost Barriers.}
We give the numerical results obtained by solving both the PDE in variational form \eqref{eq:Root-PDE} resp. \eqref{eq:Rost-PDE} as well as 
\eqref{eq:MM-Root-PDE} resp. \eqref{eq:MM-Rost-PDE} using an explicit Euler scheme. 

It is important to note that the algorithm exhibits certain challenges when dealing with 
situations where the potential associated with the time $T_k$ marginal closely approaches the initial potential $-|x - P_0|$. 

Consequently, we depict the region where these potentials remain a minimum distance of $\epsilon = 0.01$ away from the initial potential using solid lines. 
The remaining portions of the barrier are shown transparently. 
However, it is worth mentioning that we believe this boundary behavior to be an artifact of the numerical scheme and 
not indicative of the true behavior of the barriers.

\begin{figure}%[H]
\centering
\begin{subfigure}{.45\linewidth}
\centering
\includegraphics[width = 1\linewidth]{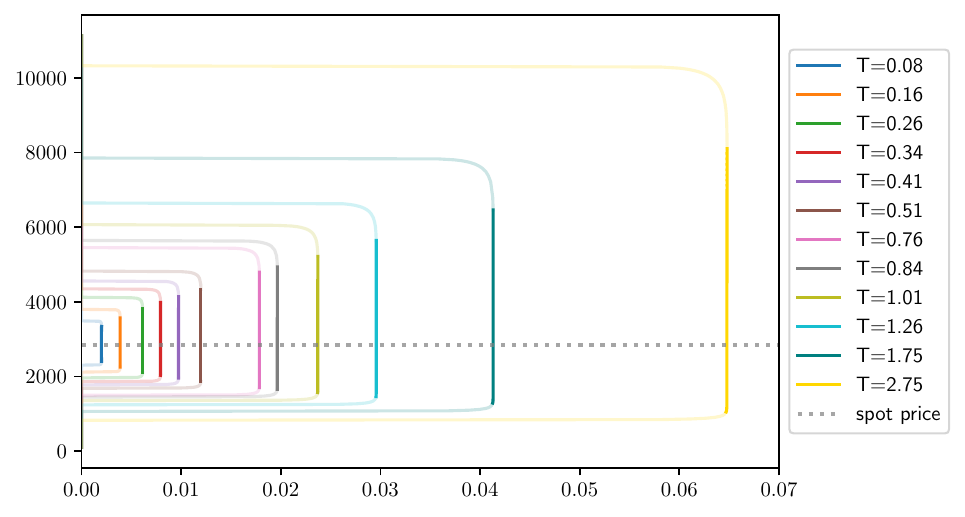}
\caption{Single marginal Root barriers}
\end{subfigure}
\quad
\begin{subfigure}{.45\linewidth}
\centering
\includegraphics[width = 1\linewidth]{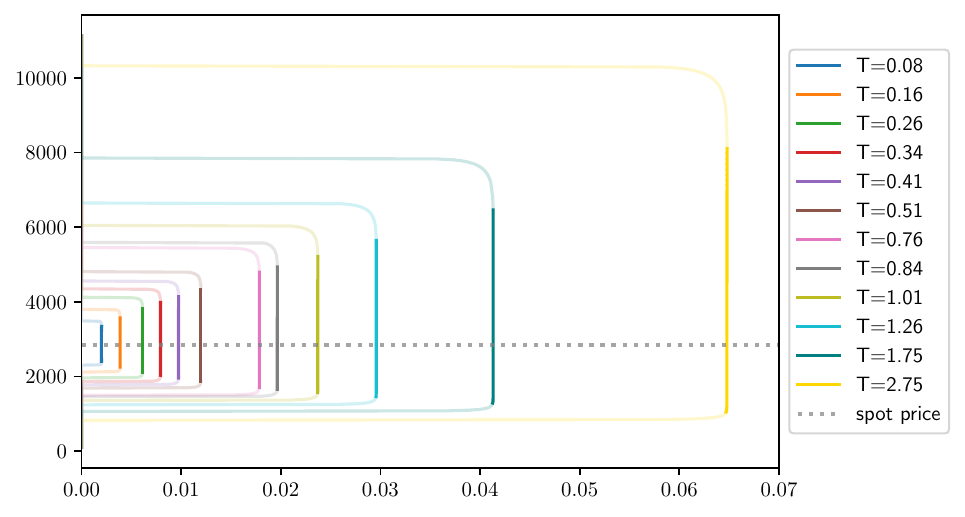}
\caption{Multi marginal Root barriers}
\end{subfigure}
\centering
\begin{subfigure}{.45\linewidth}
\centering
\includegraphics[width = 1\linewidth]{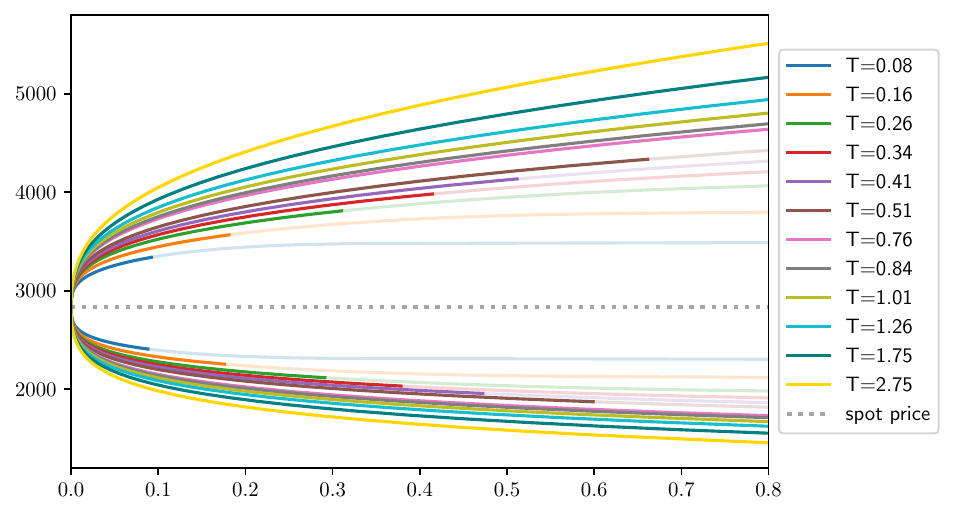}
\caption{Single marginal Rost barriers}
\end{subfigure}
\quad
\begin{subfigure}{.45\linewidth}
\centering
\includegraphics[width = 1\linewidth]{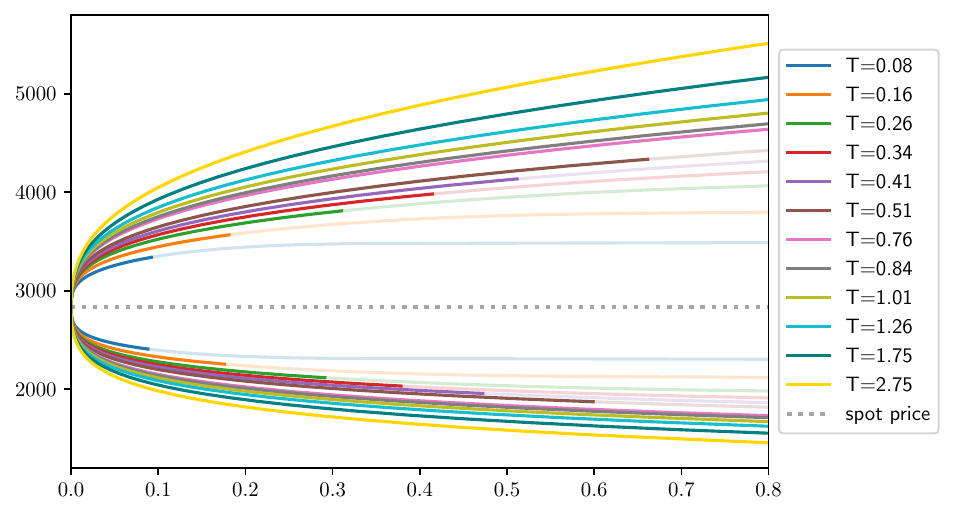}
\caption{Multi marginal Rost barriers}
\end{subfigure}
\caption{Root (top) and Rost (bottom) barriers embedding the market implied marginals of the S\&P500.}
\label{fig:BS-barriers}
\end{figure}

In the case of the Root barriers, this is very easy to see. It is straightforward to verify that the case of constant barrier functions will recover exactly the Black-Scholes model. Hence when trying to compute the barrier function corresponding to the Black-Scholes model, we would expect to recover a constant barrier. This is exactly the behaviour of the barriers shown in Figure~\ref{fig:BS-barriers}, at least in the likely region, corresponding to the `thick' lines in the figures.
%
%
%
% % % % % % % % % % % % % % % % % % % % % % % % % % % % % %  
\subsection*{Prices.}
% % % % % % % % % % % % % % % % % % % % % % % % % % % % % % 
We use the barriers obtained above to establish robust bounds for a variance call, 
as detailed in Section \ref{sec:Root-Rost} and Equation \eqref{eq:price}.
Given a maturity $T$ and a strike price $K$, our objective is to price the \emph{variance call} defined as
\[
    f(\langle \log M \rangle_T) = \left(\langle \log M \rangle_T - K \right)^+.
\]
We consider both an intermediate maturity, $T = 1.01$ as well as the terminal maturity $T = 2.75$. 
As the single-marginal and the multi-marginal barriers coincide, 
the resulting bounds are identical, regardless of whether we employ the single-marginal or multi-marginal model. 

\begin{figure}%[H]
\centering
\includegraphics[width = 1\linewidth]{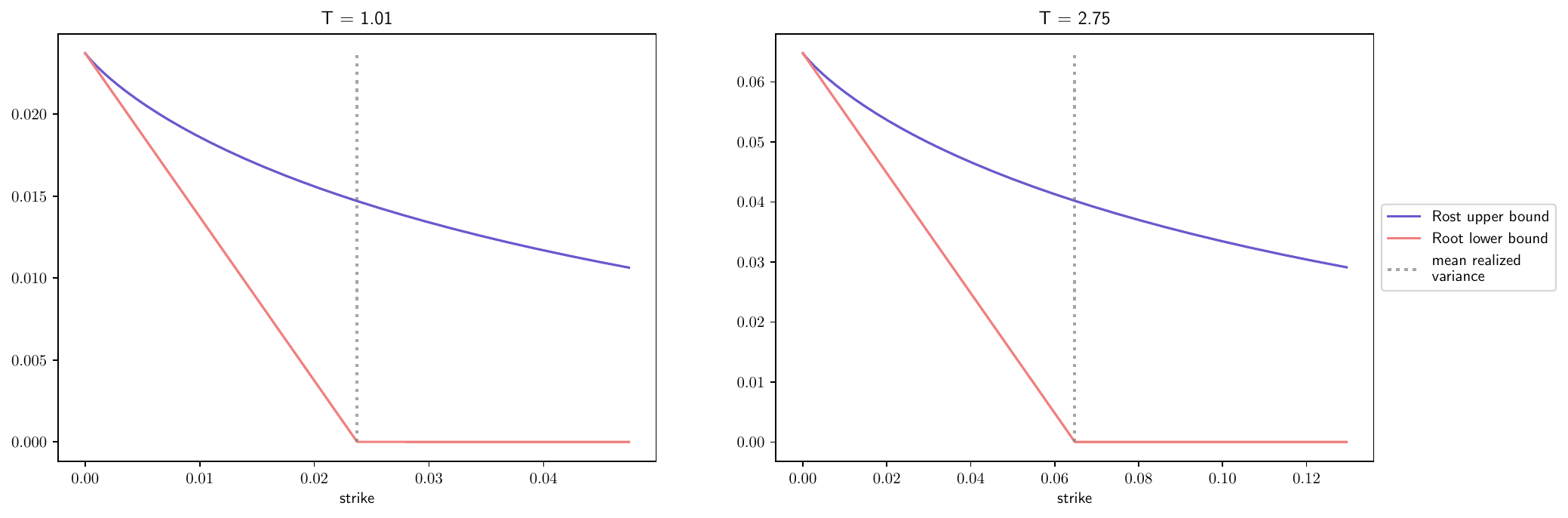}
\caption{Upper and lower robust bounds on variance call prices for two different maturities.}
\label{fig:BS-prices}
\end{figure}

In Figure \ref{fig:BS-prices} we illustrate the option prices for these variance calls for a wide range of possible strike prices.
Notably, while the upper and lower bounds coincide for the strike price 0, 
the gap between the upper and lower robust pricing bounds widens notably for increasing strike prices.
%
%
%
%
%
%
%
%
%
%
%
%
%
%
%
%
%
% % % % % % % % % % % % % % % % % % % % % % % % % % % % % %  
\newpage
\subsubsection{Heston Data} \label{sec:Heston}
% % % % % % % % % % % % % % % % % % % % % % % % % % % % % % 
\

\begin{figure}%[H]
\centering
\begin{subfigure}{0.9\linewidth}
\centering
\includegraphics[width = 1\linewidth]{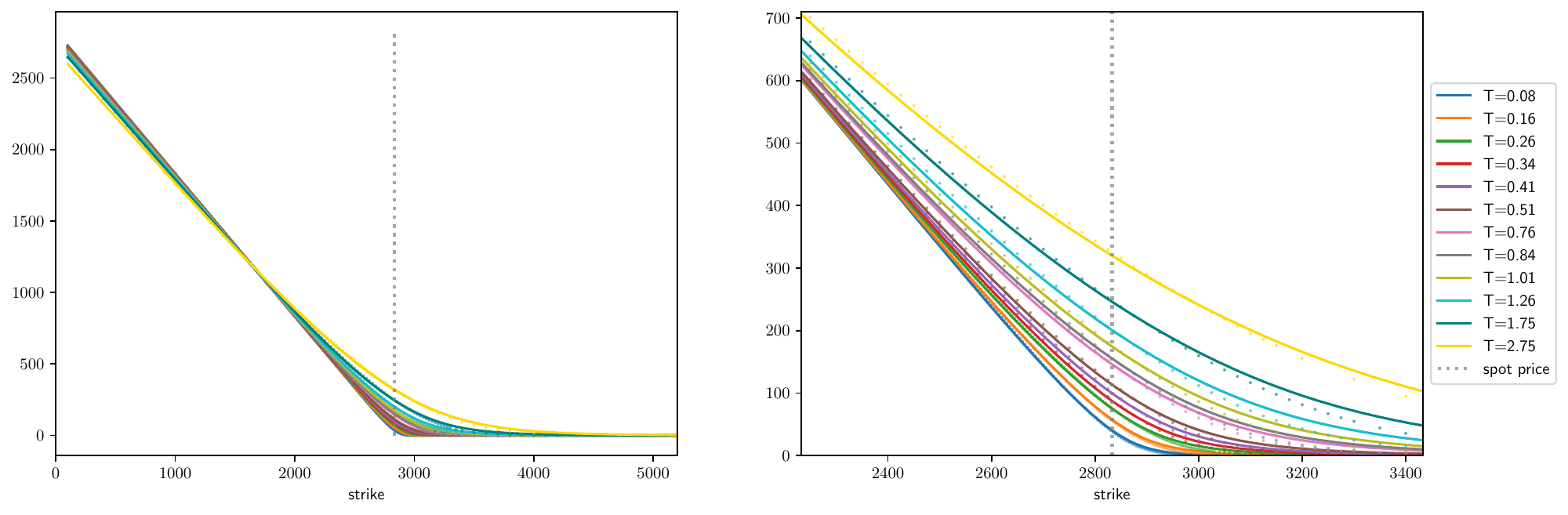}
\caption{Comparing calibrated prices to market given prices.}
\end{subfigure}
\begin{subfigure}{0.9\linewidth}
\centering
\includegraphics[width = 1\linewidth]{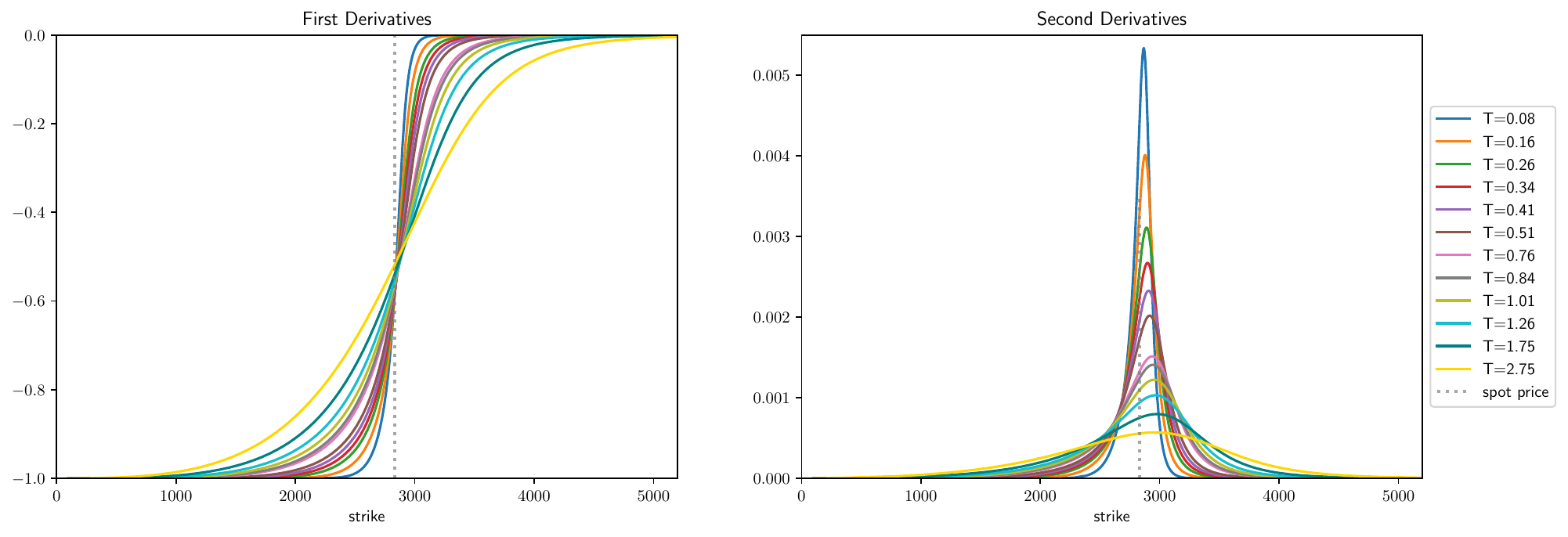}
\caption{First and second derivatives of calibrated prices.}
\end{subfigure}
\centering
\begin{subfigure}{0.9\linewidth}
\centering
\includegraphics[width = 1\linewidth]{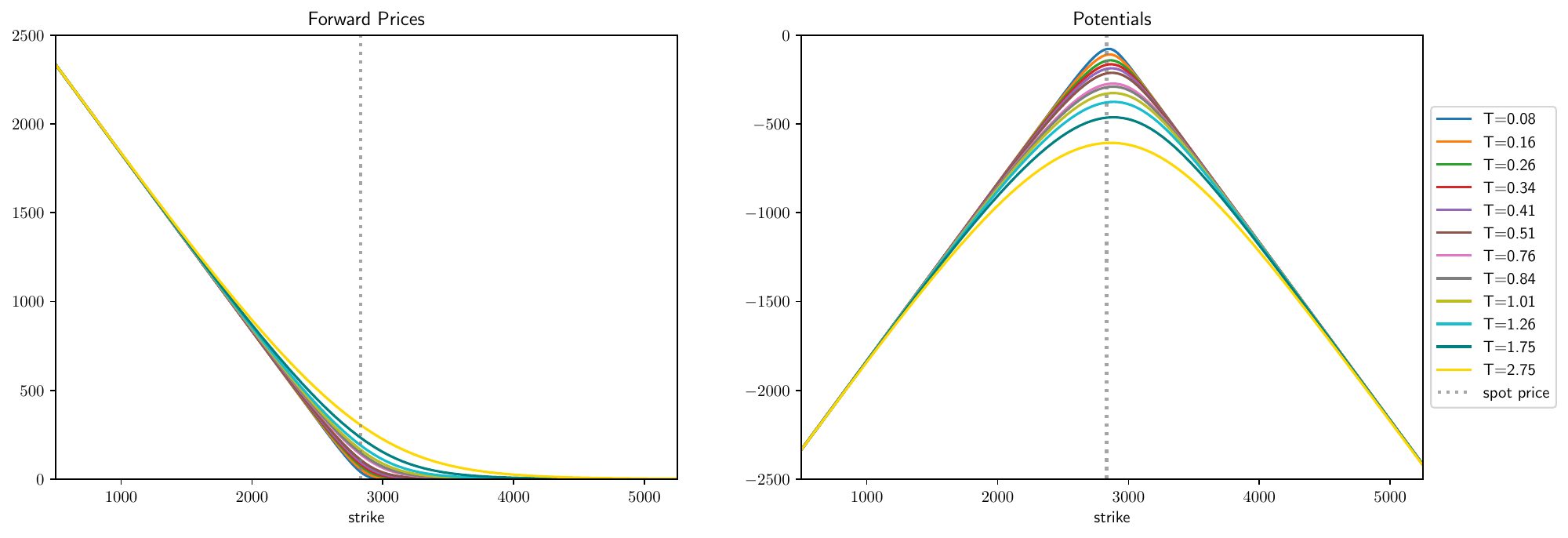}
\caption{Forward prices and the associated potentials.}
\end{subfigure}
\caption{Heston model calibration results.}
\label{fig:Heston}
\end{figure}

\noindent
\textbf{Calibration Results.} 
Calibrating a \emph{Heston model} to the market data %as described the Appendix \ref{A:Heston-calibration} 
yields the following pricing parameters.
\begin{table}[H]
\begin{center}
\begin{tabular}{ c|c|c }
    Parameter       & Calibration Result     & Parameter Meaning  
\\	\hline
    $v_0$    &  0.0158 & initial volantility  
\\  $\bar v$ &  0.0361 & long-run mean variance
\\  $\rho$   & -0.5199 & corellation of asset BM and volatility BM  
\\  $\kappa$ &  2.6280 & speed of mean reversion 
\\  $\sigma$ &  0.7902 & vol of vol
\end{tabular}
\caption{Calibration Results}
\label{tab:Heston-calibration}
\end{center}
\end{table}
The fit of the Heston prices to the market given data has improved substantially over the Black-Scholes model
as seen in Figure \ref{fig:Heston} (a), but can still not be considered ideal. 

\begin{figure}%[H]
\centering
\begin{subfigure}{0.9\linewidth}
\centering
\includegraphics[width = 1\linewidth]{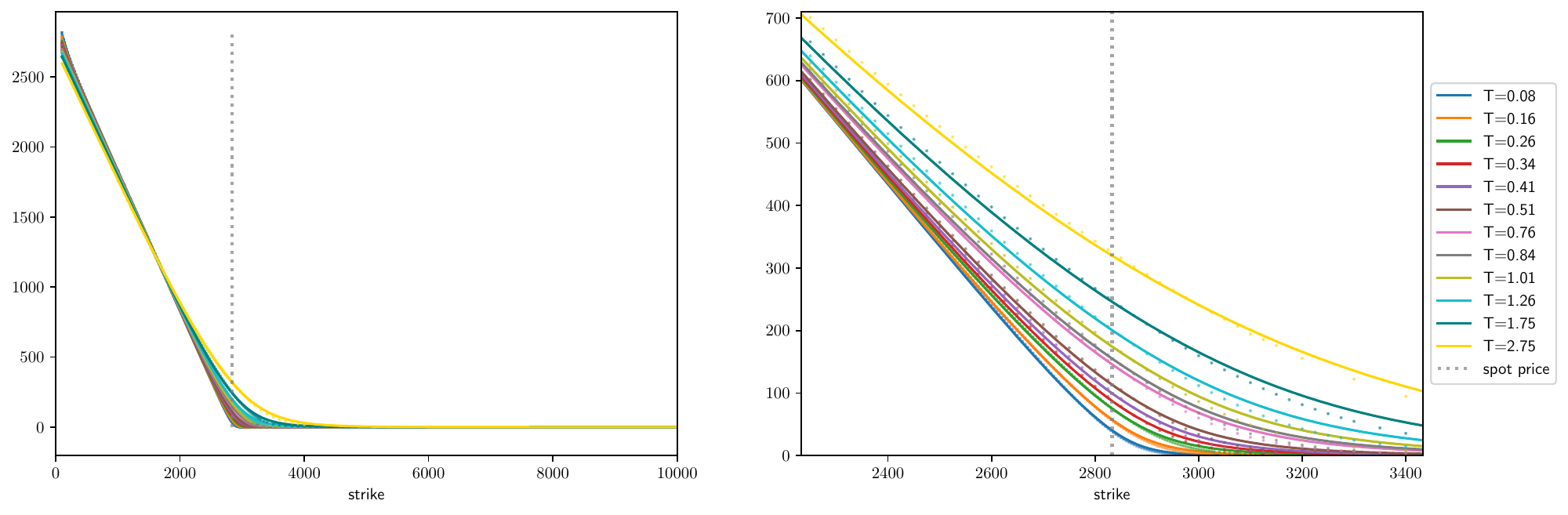}
\caption{Comparing calibrated prices to market given prices.}
\end{subfigure}
\begin{subfigure}{0.9\linewidth}
\centering
\includegraphics[width = 1\linewidth]{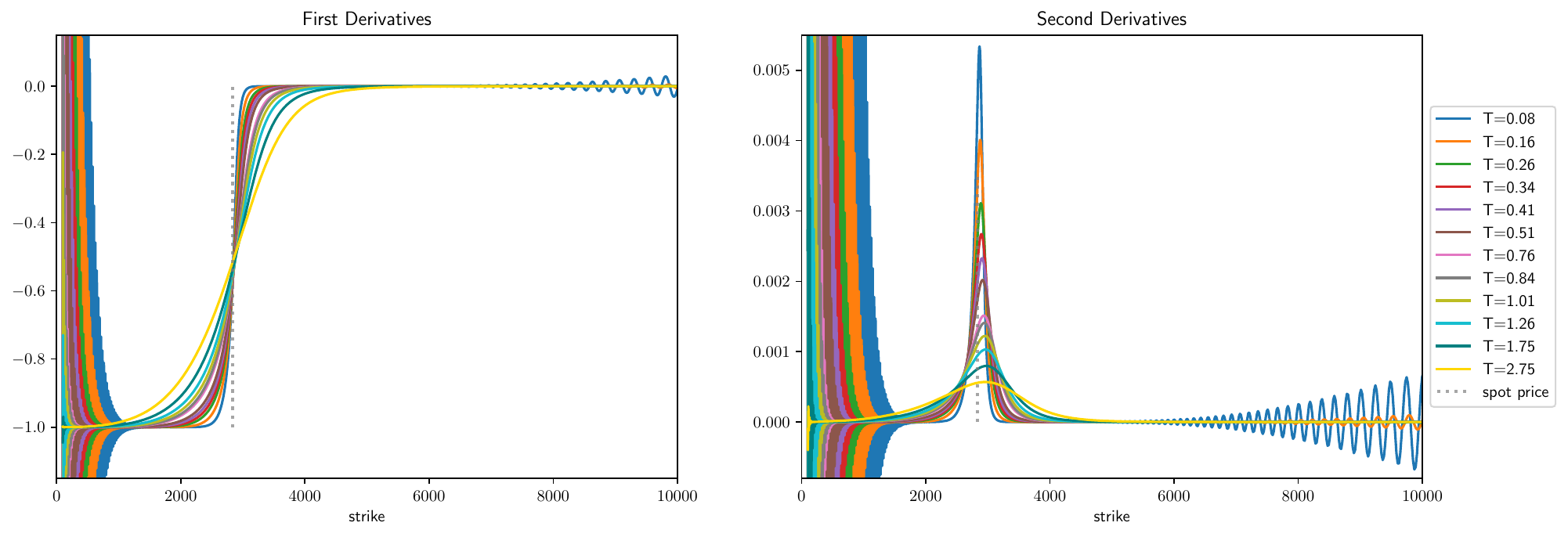}
\caption{First and second derivatives of calibrated prices.}
\end{subfigure}
\caption{Heston model calibration results before tail correction.}
\label{fig:Heston-pre-cor}
\end{figure}

However, most common implementation of the Heston pricing functionals such as the python package \verb|QuantLib| 
seem to exhibit numerical instability in the tails, 
especially prominent in the derivatives as depicted in Figure \ref{fig:Heston-pre-cor}.
To mitigate this problem it was necessary to smooth out the tails in a sensible manner. 
Just before the convexity assumption of the Heston prices breaks down, 
we replace the Heston tails with exponential tails fitted to the function values as well as the first derivatives.
This approach seems to provide a surprisingly nice fit. 

Nevertheless, this truncation of course leads to a loss of confidence in the behavior of the barriers at the boundaries regions.
We will later on depict these replaced sections with transparency to highlight the uncertainty. 

In Figure \ref{fig:Heston} (b), we present the first and second derivatives following the tail correction. 
Figure \ref{fig:Heston} (c) illustrates the associated forward prices and potentials. 


\medskip
\noindent
% % % % % % % % % % % % % % % % % % % % % % % % % % % % % %  
\textbf{Root and Rost Barriers.}
% % % % % % % % % % % % % % % % % % % % % % % % % % % % % % 
%
We give the numerical results obtained by solving both the PDE in variational form \eqref{eq:Root-PDE} resp. \eqref{eq:Rost-PDE} as well as 
\eqref{eq:MM-Root-PDE} resp. \eqref{eq:MM-Rost-PDE} using an explicit Euler scheme and the tail-corrected Heston data.

As already pointed out in the discussion of the Black-Scholes model, 
the Euler scheme does not seem to return the result we would expect when the 
potential associated with the time $T_k$ marginal too closely approaches the initial potential $-|x - P_0|$.

The algorithm does, however, seem to remain stable for a little while longer than in the Black-Scholes case,  
we depict the region where these potentials remain a minimum distance of $\epsilon = 0.0002$ away from the initial potential 
and no tail correction has taken place in solid lines. 
The remaining portions of the barrier are shown transparently. 

\medskip
\noindent
% % % % % % % % % % % % % % % % % % % % % % % % % % % % % %  
\textbf{Prices.}
% % % % % % % % % % % % % % % % % % % % % % % % % % % % % % 
We use the barriers obtained above to establish robust bounds for a variance call, 
as detailed in Section \ref{sec:Root-Rost} and Equation \eqref{eq:price}.
Given a maturity $T$ and a strike price $K$, our objective is to price the \emph{variance call} defined as
\[
    f(\langle \log M \rangle_T) = \left(\langle \log M \rangle_T - K \right)^+.
\]
We consider both an intermediate maturity, $T = 1.01$ as well as the terminal maturity $T = 2.75$. 
As the single-marginal and the multi-marginal barriers coincide, 
the resulting bounds are identical, regardless of whether we employ the single-marginal or multi-marginal model. 

\begin{figure}%[H]
\centering
\includegraphics[width = 1\linewidth]{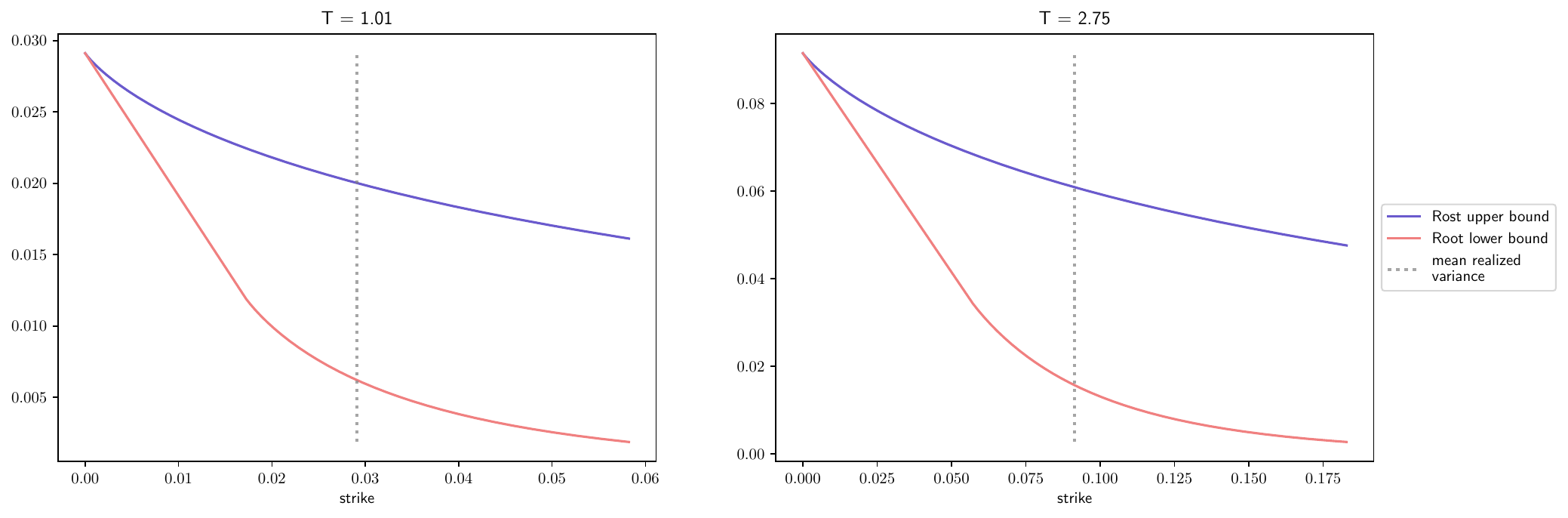}
\caption{Upper and lower robust bounds on variance call prices for two different maturities.}
\label{fig:Heston-prices}
\end{figure}

In Figure \ref{fig:Heston-prices} we illustrate the option prices for these variance calls for a wide range of possible strike prices.
Notably, similar to the Black-Scholes case, while the upper and lower bounds coincide for the strike price 0, 
the gap between the upper and lower robust pricing bounds widens notably for increasing strike prices.

Additionally, it is evident that the bounds computed here differ greatly from those computed in the Black-Scholes section, 
demonstrating a certain sensitivity of the robust bounds to variations in the data inputs.
%
%
%
%
%
%
%
%
%
%
%
%
%
%
%
% % % % % % % % % % % % % % % % % % % % % % % % % % % % % %  
\subsection{Discussion} \label{sec:discussion}
% % % % % % % % % % % % % % % % % % % % % % % % % % % % % % 

In both calibration regimes, the Root and Rost barriers adhering to the market given data which emerge from our numerical analysis appear to be in perfect consecutive order. 
In the case of the Root construction for the Black-Scholes model, this should not be surprising since (as noted above), the optimal barriers correspond to barriers which are constant in time for the time-changed model, and therefore convex ordering will imply ordering of the corresponding barriers. For the barriers based on the Heston model, it would be natural to conjecture that the barriers are ordered when the call prices come from the same parameters at different maturities (as indicated here), however it does not appear to be possible to prove this easily.

The fact that our multi-marginal barriers align precisely with their single-marginal counterparts indicates that the incorporation of earlier maturities yields no additional information for future maturities and the requirement for these barriers to be traversed chronologically imposes no additional constraints on the extremal models. As a consequence, the evidence we have collected suggests that under normal market regimes, it would be natural to assume that it would be sufficient to compute model-independent bounds based only on the terminal marginal information, and there is limited useful information contained in earlier marginals. It would be interesting to establish whether this behaviour is preserved in extreme market scenarios. 

We have carried out similar analysis on prices collected during the 2020 Covid crash. 
Although our results were similar to those presented, 
we encountered the following problems during the investigation.
Firstly, while the estimation of the option-implied interest rate yielded results somewhat aligned with the T-bill rates reported by the U.S. Department of the Treasury, the corresponding implied dividend yields deviated significantly and, notably, presented as negative. 
This suggests a potential flaw in our underlying setup for this scenario.
Furthermore, the fit of the resultant calibrated models is regrettably poor, 
rendering them unsuitable for serious consideration.
A different smoothing procedure may reveal a different answer and 
a more systematic review of abnormal market conditions may also reveal some periods where this behaviour breaks down.

The primary result from our study is that the robust pricing paradigm, when only considering the inclusion of marginal information about price-processes, do not substantially reduce price bounds beyond the information contained in the marginal law at the maturity time. 
This suggests that robust pricing is inherently limited in the extent to which the upper or lower bounds can be used for pricing, 
without the addition of further information which can reduce the class of models, 
for example through the knowledge of prices of more complex derivatives. 
Another alternative is to reduce the class of models under which pricing can happen. 
However, this approach also has potentially limited scope for reducing pricing bounds. 
Work of \cite{DoNe18} showed that provided the class of models which are included in the ``robust'' set includes some stochastic volatility model with additional noise relative to the asset, and under mild technical conditions, 
then the class of possible pricing measures is essentially the set of all calibrated measures, 
that is, we cannot significantly reduce the class $\mathcal{M}^{P_0}$ without making strong assumptions on the dynamics of the true model.

In practice, and as is well established, this means that robust methods are of limited practical application for pricing, 
a fact which has been well known for many years. 
Rather, the main applications of such methods should come through more sophisticated approaches to penalisation. 

For a completely different approach to these problems, employing modern machine learning techniques see for example 
\cite{GiSa23} using neural SDE models, while e.g.\ in \cite{NeSe23} a machine learning approch to the discrete time problem was proposed. 

Another recent methodology, often referred to as ``sensitivity analysis'', consists of restricting to models within a small ``distance'' to some reasonable reference model, 
see, e.g., \cite{BaNePa23, HeMu17, HeMuSe17} for results in continuous time and \cite{EcGuLiOb21, BdDrObWi21, ObWi21, BaWi23} for results in discrete time, as well as the references therein.

We note, however that such methods are again subject to (more subtle notions of) model-risk, and may again expose institutions adopting these metrics to unanticipated risks. 
More generally, it seems that model-risk is a phenomena which is unavoidable for accurate pricing and hedging.

\section{Conclusion}

In this paper we considered robust pricing bounds for options on variance. 
We considered the classical solutions to the robust pricing of variance options given vanilla options with the same maturity as the variance options, and (through the application of appropriate smoothing of market prices), we were able to identify the underlying geometric structure which identifies the optimal model.

Further, by incorporating market data on vanilla options with earlier maturity dates, we were able to show that, contrary to some expectations, and through explicit calculation of the geometric structure relating to the extremal models, that the extremal prices of variance options do not appear to change when we include this additional information. 
As a consequence, we conclude that there is some evidence that robust pricing bounds for variance options are not significantly changed when additional market information coming from vanilla options is incorporated. 
We conclude that there is no simple approach to eliminating model risk when pricing and hedging exotic options.
%
%
%
%
%
%
%
%
%
%
%%%%%%%%%%%%%%%%%%%%%%%%%%%%%%%%%%%%%%%%%%%%%%%%%%%%%%%%%%%
% % % % % % % % % % % % % % % % % % % % % % % % % % % % % %  
\appendix
% % % % % % % % % % % % % % % % % % % % % % % % % % % % % %
%%%%%%%%%%%%%%%%%%%%%%%%%%%%%%%%%%%%%%%%%%%%%%%%%%%%%%%%%%% 
%
%
%%%%%%%%%%%%%%%%%%%%%%%%%%%%%%%%%%%%%%%%%%%%%%%%%%%%%%%%%%%
\section{Estimating Implied Interest Rate and Dividend Yield.} \label{A:r-and-q}
%%%%%%%%%%%%%%%%%%%%%%%%%%%%%%%%%%%%%%%%%%%%%%%%%%%%%%%%%%%
In instances where interest rates and dividend yields used in pricing are unavailable, 
as is the situation with the data obtained from the CBOE Datashop, 
there exist several methods for approximating these quantities. 
A commonly used technique is utilizing the put-call parity
\begin{equation}
     \mathcal{C}(K, T) - \mathcal{P}(K, T) = P_0 e^{-qT} - Ke^{-rT}.
\end{equation}
Then taking the first derivative in the strike price $K$ yields the following formula for the \emph{implied} interest rate given $(K,T)$
\begin{align*}
 \frac{\partial}{\partial K} \left( \mathcal{C}(K, T) - \mathcal{P}(K, T)\right) 
     = - e^{-r T}
\quad \Leftrightarrow  \quad r = r(K,T) = - \frac{1}{T} \frac{\partial}{\partial K} \left(\mathcal{P}(K, T) -  \mathcal{C}(K, T)\right).
\end{align*}

Let $\mathcal{X} \subseteq [0, \infty)^2$ denote the set of market given strike-maturity pairs $(K,T)$ and let $N$ denote its cardinality. 
%Let $\mathcal{X} = \{(x^1_1, T_1),(x^1_1, T_1) , \dots,  (x^1_1, T_1) \} \subseteq [0, \infty)^2$ denote the set of market given strike-maturity pairs $(x,T)$. 
Then we estimate the \emph{implied} interest rate $\hat r$ as
\begin{equation}
    \hat r := \frac{1}{N} \sum_{(K,T) \in \mathcal{X}} r(K,T).
\end{equation}

Given an interest rate $r$ (or an estimate thereof) we estimate the \emph{implied} dividend yield given $(K,T)$ analogously 
\begin{equation}
    q(K,T) = -\frac{1}{T} \log \left( \frac{1}{P_0} 
        \left(\mathcal{C}(K, T) - \mathcal{P}(K, T) + K e^{-r T} \right)\right),
\end{equation}
hence we can define $\hat q$ by 
\begin{equation}
    \hat q := \frac{1}{N} \sum_{(K,T) \in \mathcal{X}} q(K,T).
\end{equation}
%
%
%
%%%%%%%%%%%%%%%%%%%%%%%%%%%%%%%%%%%%%%%%%%%%%%%%%%%%%%%%%%%
%%%%% REFERENCES                                        
%%%%%%%%%%%%%%%%%%%%%%%%%%%%%%%%%%%%%%%%%%%%%%%%%%%%%%%%%%%
\bibliographystyle{abbrv}
%\bibliography{../MBjointbib/joint_biblio}
\bibliography{../joint_biblio}
\end{document}